\documentclass[a4paper,12pt]{article} 
\usepackage{epsfig,amsmath} 
\usepackage{graphics}
\usepackage{ifthen}
\usepackage{amssymb}
\usepackage{rotating}

\begin{document}
\newcommand{\Pom}{I$\!$_P}                
\newcommand{\Poml}{\textrm{I$\!$P}} 
\newcommand{\etam}{\eta_{\textrm{max}}}
\def\lsim{\mathrel{\rlap{\lower4pt\hbox{\hskip1pt$\sim$}}
    \raise1pt\hbox{$<$}}}                
\def\gsim{\mathrel{\rlap{\lower4pt\hbox{\hskip1pt$\sim$}}
    \raise1pt\hbox{$>$}}}                


\def\CASCADE{{\sc Cascade}}
\def\SMALLX{{\sc Smallx}}
\def\RAPGAP{{\sc Rapgap}}
\def\PYTHIA{{\sc Pythia}}
\def\HERWIG{{\sc Herwig}}

\bibliographystyle{mysty}
\begin{flushright}
hep-ph/0402019\\
LC-PHSM-2003-100\\
LUNFD6/(NFFL-7219)2004
\end{flushright}

\begin{center}
\begin{bf}  
{\Large The Unintegrated Gluon Density in the Photon and Heavy Quark Production}\\
\end{bf}   
\vspace{15mm}  
{\large M. Hansson, H. Jung and L. J\"onsson} \\
{\it Department of Physics}\\
{\it Lund University, Lund, Sweden}\\
\end{center} 
\vspace{1.5cm}   
\begin{abstract}
The production cross section of heavy quarks in real and virtual photon-photon collisions
has been studied. The unintegrated gluon density in the photon was
obtained using the full CCFM evolution equation for the first
time. The gluon density was implemented in the Monte Carlo generator
CASCADE, and cross sections for heavy quark production in $e^+e^-$
collisions were calculated and compared to LEP data. Also, predictions
for heavy quark cross sections in $e^+e^-$ and $\gamma\gamma$
collisions at TESLA energies are given.
\end{abstract}
 
\section{Introduction}
The production of heavy flavour ($b$ and $c$ quarks) in $\gamma
\gamma$ collisions has been studied at
LEP~\cite{LEP1, LEP2, LEP3, LEP4, LEP5, LEP6, LEP7, LEP8, LEP9, LEP10} for many years and will be
an important field of study at TESLA~\cite{tesla, Jankowski}. It has also been studied in
proton-photon collisions at HERA~\cite{H11, H12, H13, Zeus1, Zeus2, Zeus3}, and in proton-proton collisions
at the TEVATRON~\cite{Tevatron1, Tevatron2, Tevatron3,
  Tevatron4}. Whereas originally the
measured charm cross sections were well described by the parton
evolution according to the standard collinear
approach, the measured beauty cross sections
were a factor 2-4 larger than predictions. It has been shown
\cite{Cacciari:2002pa,Cacciari:2003uh,Gardi:2003ar} that an improved description of
the heavy quark fragmentation function together with new
parametrizations of structure functions and resumming next-to-leading
logarithms in NLO QCD calculations
result in an improved description of the B meson cross section in
$p\bar{p}$ and $e^{+}e^{-}$ collisions. Also the application of
$k_{t}$-factorization and an unintegrated gluon density for the 
proton gives beauty cross sections which are in good agreement with data
from $p\bar{p}$ collisions~\cite{ccfmtevatron1, ccfmtevatron2}. The aim of
this paper is to investigate if the $b\bar{b}$ cross section also can
be explained in $\gamma \gamma$ collisions by using $k_{t}$-
factorization and an unintegrated gluon density for the photon. 

In a previous study \cite{Motyka}, using $k_{t}$-factorization, the unintegrated
gluon density in the photon was calculated using a simplified solution
to the CCFM formalism \cite{ccfm1,ccfm2,ccfm3,ccfm4} proposed by \cite{kmr}.
As an alternative approach also a generalization of the GBW saturation
model \cite{Timneanu,gbw1} was used to obtain the unintegrated gluon
density. In the present study we will  use the complete CCFM evolution
equations to determine the unintegrated gluon density in the photon and
calculate the heavy quark cross section with help of the CASCADE Monte
Carlo program. 

\section{The structure of the photon}
\label{photonstructure}
\begin{figure}
\begin{center}
\resizebox{12cm}{!}{\includegraphics{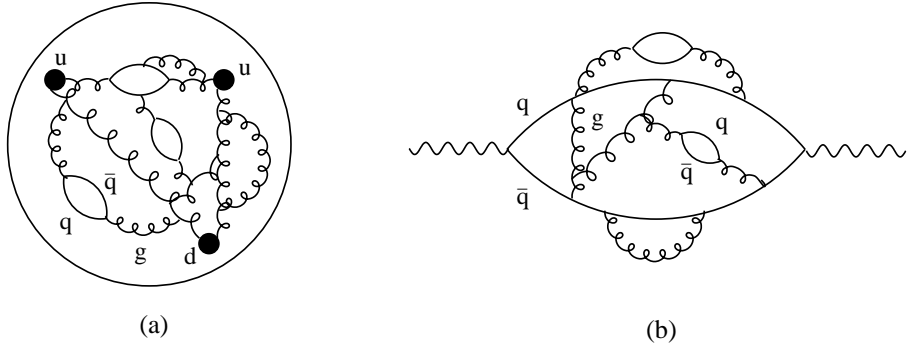}}
\caption{\it The structure of the proton (a) and the photon (b).}  
\label{photon}
\end{center}
\end{figure}

It is well known that the proton not only consists of three valence
quarks, but also of virtual gluons that hold the quarks
together. These gluons can split up in quark-antiquark pairs, called
sea quarks, which in turn can emit gluons, see Figure~\ref{photon}a. A
photon colliding with a proton may therefore interact either with one
of the valence quarks or with one of the sea quarks.

In analogy with the proton, a photon can be seen as a flux of (virtual)
quarks and gluons. However, the photon does not consist of any valence
quarks, and must first fluctuate into a virtual quark-antiquark pair
which can emit gluons, see Figure~\ref{photon}b. When a photon
interacts with a virtual quark in another photon, the former is said
to resolve the latter. How much of the photon that is resolved depends
on the resolving power (virtuality $Q^{2}$) of the photon. So, what a photon with a low $Q^{2}$ will resolve as a
quark, a photon with a higher $Q^{2}$ might resolve as a quark and a
gluon which is radiated by the
quark, sharing the original quark momentum. A photon with a very high $Q^{2}$ may also resolve the soft
gluon as having split up in a quark-antiquark pair. This means that the more
we resolve the more quarks we see at smaller $x$, since each daughter
particle must have less momentum than the mother. The density of
quarks and gluons thus depends on $Q^{2}$, a phenomenon
called scaling violation. For photons, this $Q^{2}$ dependence occurs
already with the anomalous part of the splitting, i.e. the splitting
of the photon into a quark-antiquark pair. The density of quarks and gluons is described by the density function
$f_{i}(x,\mu ^{2})$ which is just the probability that a parton of
type $i$ (quark or gluon) is carrying a longitudinal momentum fraction
$x$ of the photon, at a scale $\mu ^{2}=Q^{2}$. The scale is generally
denoted $\mu ^{2}$, since it does not always have to be the virtuality
of the photon that sets the scale. The structure function of the photon can
then be written as \[F_{2}^{\gamma}(x,\mu ^{2})=x\sum
_{i}e_{i}^{2}f_{i}^{\gamma}(x,\mu^{2})\] where $e_{i}$ is the electric charge of
parton $i$ and one sums over all quark flavours. Of course, the photons cannot interact with the gluons,
since $e_{gluon}=0$, and $F_{2}^{\gamma}$ does not depend on the gluon
structure function explicitly (in LO and NLO in the DIS factorization scheme \cite{Brock}). However, since the quark density at
small $x$ is driven by the gluon density, it is possible to extract
$f_{gluon}(x,\mu ^{2})$.
In general, the cross section that a parton $i$ from one photon will
interact with a  parton $j$ from another photon and form the final
state $X$ is 
\begin{equation}
\sigma =\sum _{i,j}\int \frac{dx_{1}}{x_{1}}\frac{dx_{2}}{x_{2}}\cdot
f_{i}^{\gamma}(x_{1},\mu ^{2})\cdot
f_{j}^{\gamma}(x_{2},\mu ^{2})\cdot \hat{\sigma}_{i+j\rightarrow X}(x_{1},x_{2},s) 
\label{xsec}
\end{equation}
where the parton densities $f_{i}^{\gamma}$ and $f_{j}^{\gamma}$ describe the probability of finding
partons $i$ and
$j$ with momentum fractions $x_{1}$ and $x_{2}$ and the partonic cross section
$\hat{\sigma}$ describes the probability that the two particles $i$
and $j$ will create the final state $X$. The partonic cross section
is proportional to the square of a Matrix Element (ME), which in turn
depends on the masses, virtualities, couplings etc. of the particles
involved. The
integral is over all $x_{1}$ and $x_{2}$ allowed by the
kinematics. Unlike the partonic cross section, the parton densities cannot be calculated from first principles. Instead,
experimental fits and phenomenological methods must be used in order
to describe the parton distributions inside the photon. 

\section{$k_{t}$-factorization}
\label{ktfactor}
All factors in equation~(\ref{xsec}) above only depend on the
longitudinal momentum fraction $x$ and the scale $\mu ^{2}$.
This means that all partons are assumed to travel in the
same direction as the incoming particle, i.e. they have no transverse
momenta. It is therefore called a collinear approximation. However, at
large energies, where small $x$ gluons are probed, the transverse momenta $k_{t}$ of the partons are
expected to be important. 
Therefore, the cross sections are factorized~\cite{catani, ktfac2, ktfac3, ktfac4} into a
$k_{t}$-dependent partonic cross section
$\hat{\sigma}(x,k_{t}^{2},\mu ^{2})$, where the incoming partons are
treated off-mass shell, and a
$k_{t}$-dependent parton density function
$\mathcal{A}(x,k_{t}^{2},\mu ^{2})$. The equivalent to
equation~(\ref{xsec}) then becomes 
\begin{equation}
\sigma =\sum _{i,j} \int
\frac{dx_{1}}{x_{1}}\frac{dx_{2}}{x_{2}}dk_{t1}^{2}dk_{t2}^{2}\mathcal{A}_{i}^{\gamma}(x_{1},k_{t1}^{2},\mu^{2})\mathcal{A}_{j}^{\gamma}(x_{2},k_{t2}^{2},\mu^{2})\hat{\sigma}_{i+j\rightarrow
X}(k_{t1},k_{t2},x_{1},x_{2},s).
\label{xsec2}
\end{equation}
The function $\mathcal{A}(x,k_{t}^{2},\mu ^{2})$ is called the ($k_{t}$-)
unintegrated gluon density and is related to the collinear one by
\begin{equation}
g(x,\mu^{2})\simeq
\int_{0}^{\mu^{2}}dk_{t}^{2}\mathcal{A}(x,k_{t}^{2},\mu^{2}),
\label{pdf1}
\end{equation}
where $\mathcal{A}(x,k_{t}^{2},\mu ^{2})$ describes the
probability to find a gluon with a longitudinal momentum fraction $x$
and a transverse momentum $k_{t}$ at a scale $\mu ^{2}$.
Inverting the relation (\ref{pdf1}) gives that an unintegrated
gluon density may be obtained by differentiating an integrated gluon
density with respect to $\mu ^{2}$,
\begin{equation}
\mathcal{G}(x,k_{t}^{2})\equiv \left. \frac{dg(x, \mu ^{2})}{d\mu ^{2}} \right|_{\mu ^{2}=k_{t}^{2}}.
\label{pdf2}
\end{equation}
Note that the unintegrated gluon density obtained in this way only
depends on one scale, since $\mu ^{2}=k_{t}^{2}$. To distinguish the
different parton distributions, the notation following \cite{Andersson} is used:
$g(x,\mu ^{2})$ is the parton distribution in the collinear approach,
while $\mathcal{G}(x,k_{t}^{2})$ and
$\mathcal{A}(x,k_{t}^{2},\mu^{2})$ are the one- and two-scale distributions in the
$k_{t}$-factorization approach.
When using off-shell matrix elements of leading order
($O(\alpha _{s})$), some next-to leading order ($O(\alpha _{s}^{2})$)
processes in the collinear approach are effectively included, in
addition to all collinear leading order processes.

\section{Heavy quark production processes}
\begin{figure}
\begin{center}
\resizebox{12cm}{!}{\includegraphics{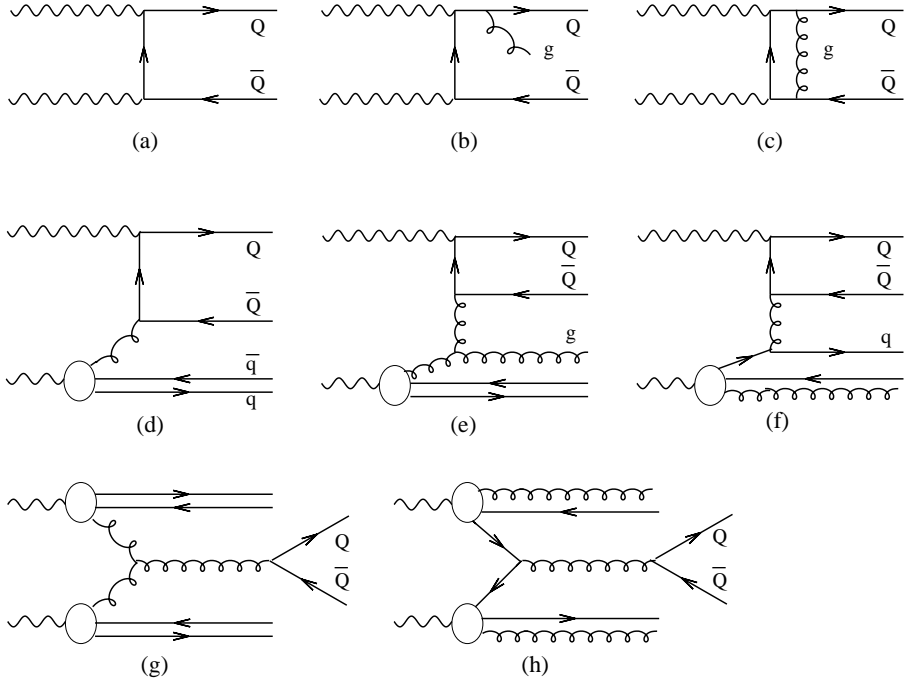}}
\caption{\it Production mechanisms in heavy quark production.}  
\label{prodmech}
\end{center}
\end{figure}

Heavy quark production in photon collisions can be explained by three
mechanisms, shown in Figure~\ref{prodmech}. At small $\sqrt{s}$, the
cross section is dominated by the direct part (a-c). In this case, the
photons couple directly to the quarks, and the process is therefore
not dependent on the quark and gluon content in the photon. Hence, the
lowest, or zeroth, order process (a) can be calculated in pure QED, while
first order corrections, including real (b) and virtual (c) gluon
radiation, also depend on the QCD coupling constant. These
corrections have been calculated in \cite{Drees} and were found to contribute
30$\%$ of the cross section in the direct case. It has also been shown
that the direct part alone (including corrections) cannot reproduce
the measurements of $b$ and $c$ quark production in $e^{+}e^{-}$, $ep$
and $pp$ collisions. 

In the single resolved case (d-f), one of the photons splits up in a flux
of quarks and gluons, where one of these gluons fuse with the
other photon. The next-to-leading order corrections are shown in (e)
and (f). Since this process depends on the quark and gluon distribution
in the photon, it cannot be calculated using perturbative QCD, but
must be treated phenomenologically. The direct and single resolved
processes are predicted to give equal contributions to the cross
section at $\sqrt{s} \approx 200$~GeV \cite{Drees}. In (g) and (h) is
shown the double resolved case, where both photons split up and create
a heavy quark pair. This process gives a smaller contribution than the
direct and single resolved processes, since it depends on $\alpha
^{2}\approx 0.014$, but it becomes increasingly important with
increasing energy. 

In $e^{+}e^{-}$ reactions there are additional contributions to heavy
quark production, which have been discussed in \cite{Boos}. The most
important contribution comes from initial or final state radiation of
a $\gamma ^{\ast}/Z$, which decays into a heavy quark pair. The cross
section of other processes discussed in \cite{Boos} are typically
several orders of magnitude smaller at linear collider energies. Only
the diagrams shown in Figure~\ref{prodmech}a, d, e and g have been taken into
account in this analysis.

\section{Higher order processes}
\begin{figure}
\begin{center}
\resizebox{5cm}{!}{\includegraphics{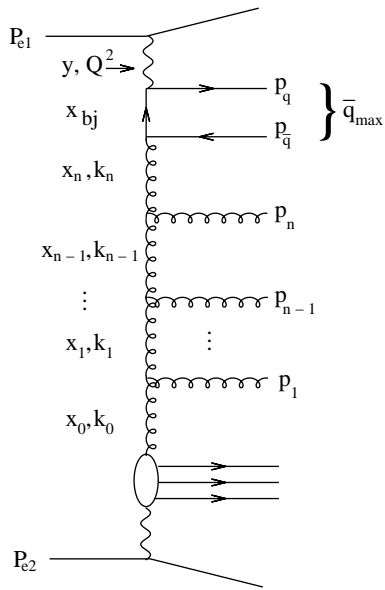}}
\caption{\it Parton ladder created by gluon emissions.}  
\label{ladder}
\end{center}
\end{figure}

The processes in Figure~\ref{prodmech} are all of zeroth, first or
second order in the strong coupling $\alpha _{s}$, and the matrix
elements have all been calculated using perturbative QCD. However, in
high energy collisions it is possible to have many gluon emissions,
making the zeroth, first and even second order calculations
insufficient to decribe the data. Due to the self coupling of the
gluons, the complexity of the processes increases rapidly with
increasing order, making perturbative calculations very
difficult. Instead one uses evolution equations to generate the higher
order emissions. These equations describe, under certain
approximations, how a mother parton is split into two daughter
partons, one of which is emitted whereas the other continues as a
propagator parton. The daughter propagator parton is then split into
two daughters and so on. This creates a parton ladder, as illustrated in Figure~\ref{ladder}. The general structure of an evolution equation is \[f_{j}=f_{0}+\int f_{i}\cdot P_{i\rightarrow j,k}\]
where $f_{i}$ and $f_{j}$ are the parton density functions for the
mother and daughter propagator gluons, $f_{0}$ is the input gluon
density function and $P_{i\rightarrow j,k}$ is the splitting kernel describing
the probability that the parton $i$ is split into two partons $j$ and
$k$. In the following discussion, the evolution starts at the bottom in Figure~\ref{ladder} and
continues towards the quark box. The possible splittings are
$q\rightarrow qg$, $g\rightarrow q\bar{q}$ and $g\rightarrow
gg$. There are a number of different evolution equations, each one
taking different parts of the full calculation into
account and evolving the density functions in different variables. A
short presentation of three different approaches follows, where the
first is valid for large $\mu ^{2}$, the second is valid for small $x$
and moderate $\mu ^{2}$, and the third is valid for both these regions. The latter is of special importance for this paper.   

\subsection{The DGLAP evolution}
The DGLAP~\cite{dglap1,dglap2,dglap3,dglap4} evolution equation is of the form
\begin{equation}
\frac{df_{j}(x,\mu^{2})}{d \ln \mu^{2}} = \frac{\alpha _{s}(\mu^{2})}{2 \pi} \sum_{i}
\int_{x}^{1} \frac{dx'}{x'}f_{i}(x',\mu^{2})P_{i\rightarrow j,k}(z)
\label{dglap}
\end{equation}
where $f(x,\mu^{2})$ is the density of partons carrying a longitudinal
momentum fraction $x$ probed at a scale $\mu^{2}$, and
$P_{i\rightarrow j,k}(z)$ describes the probability that a parton $i$
is split into two partons $j$ and $k$ with a fraction $z=\frac{x}{x'}$
and $1-z$ of the
original parton momentum, respectively. The probability that a gluon
splits into two gluons is given by
\begin{equation}
P_{g\rightarrow g,g}(z)= \frac{1}{1-z}-2+z(1-z)+\frac{1}{z}
\label{dglapsplitting}
\end{equation}
where the terms $\frac{1}{1-z}$ and $\frac{1}{z}$ are called the
singular terms, since they give infinite
contributions when $z\rightarrow 1$ and $z\rightarrow 0$,
respectively.
The equation (\ref{dglap}) thus
describes the probability change of finding a parton of type $j$ with
momentum fraction $x$ as we increase the scale $\mu^{2}$. In the
DGLAP formalism, the gluon chain in Figure~\ref{ladder} is assumed to
be strongly ordered in virtuality
\begin{equation}  
\mu^{2}\gg |k_{n}^{2}|\ldots\ \gg |k_{1}^{2}|\gg |k_{0}^{2}|,
\label{virtorder}
\end{equation}
 where $k_{i}$ is the four momentum of parton $i$. 
This means that in each splitting $i\rightarrow i+1,j$ one can approximate
$k_{0}^{2}=k_{1}^{2}=\ldots\ k_{i}^{2}=0$ compared to $k_{i+1}^{2}$, and since
$k_{i}^{2}=m_{i}^{2}$ these partons are considered to be massless (or
on-shell). It can be shown that the ordering in virtuality implies that also the
transverse momentum of the propagator partons are strongly ordered (at
small $z$) according to
\begin{equation}
\mu^{2}\gg |k_{tn}^{2}|\ldots\ \gg |k_{t1}^{2}|\gg |k_{t0}^{2}|
\label{DGLAPkt}
\end{equation}
where $k_{ti}=(0,k_{xi},k_{yi},0)$. Hence, in each splitting one can approximate
$k_{t0}^{2}=k_{t1}^{2}=\ldots\ k_{ti}^{2}=0$, which means that the
DGLAP approach is a collinear approximation (which is obvious since
there is no $k_{t}$-dependence in (\ref{dglap})). It can be shown, that using the DGLAP evolution is equivalent to resum terms of the form $(\alpha_{s}\ln(\mu^{2}))^{n}$ in the expansion of the cross section. Hence, the DGLAP approximation is only valid at large $\mu^{2}$ where these terms will dominate.

\subsection{The BFKL evolution}
The BFKL~\cite{bfkl1,bfkl2,bfkl3} evolution equation resums the terms
$(\alpha_{s} \ln(\frac{1}{x}))^{n}$ in the expansion, and is thus only
valid at small $x$. It is of the form  
\begin{equation}
\frac{d\mathcal{G}(x,k_{t}^{2})}{d \ln( \frac{1}{x})}=\int
dk_{t}^{'2}\mathcal{G}(x,k_{t}^{'2})\cdot K(k_{t}^{2},k_{t}^{'2}).
\label{bfkl}
\end{equation}
Here, the function $K$ is the splitting kernel equivalent to $P$ in
(\ref{dglap}). The evolution is made in increasing $\ln(
\frac{1}{x})$, since
\begin{equation}
x_{0}^{2}\gg x_{1}^{2} \ldots\ \gg x_{n}^{2}\gg x_{Bj}^{2} 
\label{bfkl2}
\end{equation}
has been assumed. This implies that the emitted gluons will take a large fraction of the propagator momentum.
However, there is no ordering in $k^{2}$ or $k_{t}^{2}$, so the
collinear approximation can not be used, and the incoming partons of
the matrix elements must
be taken off-shell (the particles can have a virtual mass). Another important consequence of BFKL is that unintegrated parton densities must be used in (\ref{bfkl}), i.e. they must depend on $k_{t}$.

\subsection{The CCFM evolution}
\begin{figure}
\begin{center}
\resizebox{12cm}{!}{\includegraphics{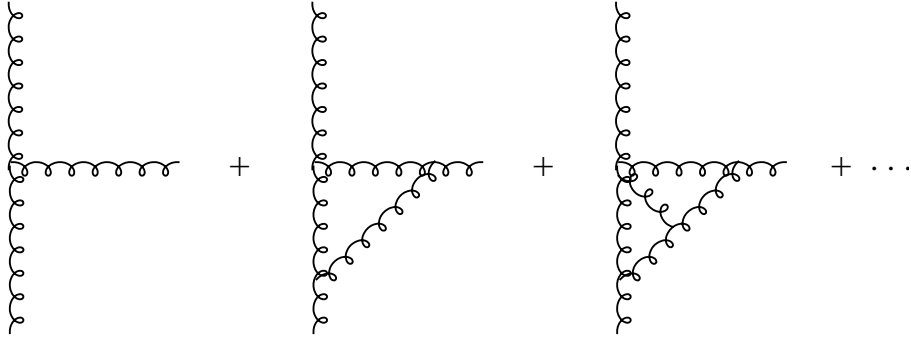}}
\caption{\it The Reggeization of the gluon vertex.}  
\label{Regge}
\end{center}
\end{figure}

The CCFM \cite{ccfm1,ccfm2,ccfm3,ccfm4} evolution equation is valid
both at large and small $x$, since it resums terms of both the form
$(\alpha_{s} \ln (\frac{1}{x}))^{n}$ and $(\alpha_{s} \ln
(\frac{1}{1-x}))^{n}$. This means that at large $x$ the CCFM evolution
will be DGLAP-like, and at small $x$ it will be BFKL-like. The CCFM
evolution includes angular ordering in the initial state cascade,
which means that the emission angles of the partons with respect to
the propagator increases as one moves towards the quark box, 
\begin{equation}
\Xi \gg \xi _{n} \gg \ldots\ \xi _{1} \gg \xi _{0}, 
\label{angord}
\end{equation}
where the maximum allowed angle $\Xi$ is set by the hard quark box,
\[p_{q}+p_{\bar{q}}=\Upsilon (P_{e1}+\Xi P_{e2})+\vec{Q}_{t}.\] This
is written in the Sudakov variables, where $p_{q}$, $p_{\bar{q}}$,
$P_{e1}$ and $P_{e2}$ are the four momenta of the produced heavy quarks
and the incoming particles, respectively (see Figure~\ref{ladder}), $\Upsilon$ and $\Upsilon
\Xi$ are the positive and negative light-cone momentum fractions of
the heavy quark pair, and $\vec{Q}_{t}$ is the sum of the transverse
momentum vectors of the heavy quark pair. The momenta of the emitted gluons
can be written similary, 
\begin{equation}
p_{i}=\upsilon_{i}(P_{e1}+\xi _{i}
P_{e2})+p_{ti}, \  \xi _{i}=\frac{p_{ti}^{2}}{s\upsilon
_{i}^{2}},
\label{emang1}
\end{equation}
where $\upsilon _{i}=(x_{i-1}-x_{i})$ is the momentum
fraction of the emitted gluon, $p_{t}$ is the
transverse momentum of the gluon, and $s=(P_{e1}+P_{e2})^{2}$ is, as
usual, the squared center of mass energy. Here, we have assumed that
all particles are massless. The CCFM equation is written as
\[\bar{q}^{2}\frac{d}{d\bar{q}^{2}}
\frac{x\mathcal{A}(x,k_{t}^{2},\bar{q}^{2})}{\Delta
_{s}(\bar{q}^{2},\mu _{0}^{2})}=\int dz
\frac{d\phi}{2\pi}\frac{\tilde{P}(z,k_{t}^{2},(\bar{q}/z)^{2})}{\Delta
_{s}(\bar{q}^{2},\mu _{0}^{2})}x'\mathcal{A}(x',k_{t}^{'2},(\bar{q}/z)^{2})\]
where 
\begin{equation}
\bar{q}_{i}=\frac{p_{ti}}{1-z_{i}}=x_{i-1}\sqrt{s\xi
_{i}}
\label{emang2}
\end{equation}
is the rescaled transverse momenta of the emitted gluons and
$z_{i}=\frac{x_{i}}{x_{i-1}}$. In this formalism, (\ref{angord})
becomes 
\begin{equation}
q_{i}>z_{i-1}q_{i-1}.
\label{qcond}
\end{equation}
When $z\rightarrow 1$ we have $q_{i}>q_{i-1}$, i.e. ordering in rescaled
transverse momentum, which means that the evolution is
DGLAP-like. In the limit $z\rightarrow 0$ the angular ordering gives
no restrictions on the rescaled transverse momentum. Also, (\ref{bfkl2}) holds
because of the definition of $z$. This means that the evolution is BFKL-like.  
The Sudakov form factor $\Delta _{s}$ describes the
probability that there are no emissions from the starting scale $\mu
_{0}^{2}$ to the maximum rescaled transverse momentum $\bar{q}^{2}_{max}$.
The CCFM splitting
function $\tilde{P}$ is defined as
\begin{equation}
\tilde{P}_{g}(z,k_{t}^{2},(\bar{q}/z)^{2})=\frac{\bar{\alpha}_{s}(q^{2}_{i}(1-z_{i})^{2})}{1-z_{i}}+\frac{\bar{\alpha}_{s}(k_{ti}^{2})
}{z_{i}}\Delta _{ns}(z_{i},k_{ti}^{2},q_{i}^{2}), 
\label{ccfmsplitting}
\end{equation}
which is somewhat different than the DGLAP splitting function
(\ref{dglapsplitting}). First of all, the CCFM splitting function only includes the singular parts of the
DGLAP splitting function. The other
difference is that there is one additional function
$\Delta _{ns}$, called the non-Sudakov form factor, in
(\ref{ccfmsplitting}). The non-Sudakov form factor originates from the
fact that, in CCFM and BFKL, all virtual corrections in the gluon vertex are
automatically taken into account, see
Figure~\ref{Regge}. This is called the Reggeization of the gluon vertex.

\section{Monte Carlo simulations}
As was already mentioned, higher order processes are very difficult to
calculate using perturbative QCD. Moreover, it is not quarks that are
detected in the experiments, but hadrons. The transition from parton
level (Figure~\ref{prodmech}, for example) to hadron level takes place
over long distances where $\alpha _{s}$ is large. Hence, perturbation
theory is not valid here. To overcome this problem, Monte Carlo (MC)
generators are used. For each event, these programs generate all the
particles, their four-momenta and all the kinematic variables
according to certain theoretical prescriptions. The higher order gluon
emissions are simulated by evolving the parton ladder according to
some evolution equation (DGLAP, BFKL, CCFM etc.), and the transitions
to hadron level can also be implemented via hadronization models.


\subsection{CASCADE}
The Monte Carlo generator CASCADE~\cite{cascade} is based on the CCFM
evolution equation and thus uses unintegrated parton desities and
off-shell matrix elements. For technical reasons, a backward evolution
is used, where first the hard scattering is generated and then the
initial state cascade (the gluon ladder) is evolved from the quark box to the incoming particles. In this
evolution only gluon emissions are treated, that is, only the
splitting $g\rightarrow gg$ is considered. CASCADE also performs the
hadronization, using the Lund string model in JETSET/PYTHIA~\cite{jetset}. 

\section{The unintegrated gluon density}
\label{mygluon}
The evolution machinery was first tested using the one-loop
(DGLAP-like) evolution for the proton and the photon. Then, the
unintegrated gluon density for the photon was obtained using the full
CCFM evolution. The uncertainties in the evolution are discussed in section~\ref{unc1}.
Since the distributions obtained with the CCFM evolution were compared with the derivative of
standard gluon density functions, taken from~\cite{pdflib}, a short discussion of the
differentiation method will be given first. Alternative methods to
calculate the unintegrated gluon density can be found in \cite{Motyka,Shoshi:2002fq}.

\subsection{Numerical differentiation method}
\begin{figure}
\begin{center}
\resizebox{14cm}{!}{\includegraphics{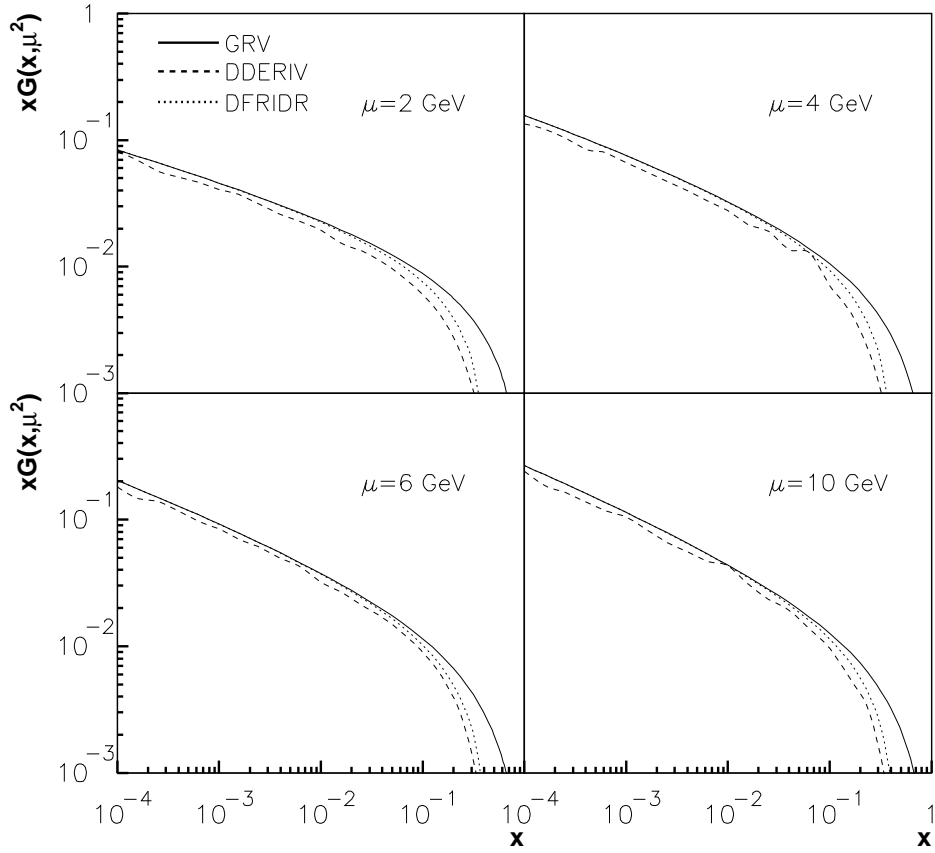}}
\caption{\it The GRV distribution differentiated with DFRIDR and DDERIV, and then integrated, compared with the original GRV distribution.}  
\label{deriv2}
\end{center}
\end{figure}

Two numerical differentiation methods were tested: DDERIV~\cite{dderiv} and
DFRIDR~\cite{dfridr}, which are based on Rombergs principle of sequence
extrapolation and Richardson's deferred approach to the limit, respectively. 
The first test was to differentiate the simple function
$f(x)=-\frac{1}{x}$ with respect to $x$. Here, both methods gave good results, but only after adjusting a parameter in DDERIV.
A second test was to differentiate the GRV\footnote{GRV-G HO (GRV for
 the photon) was used.}~\cite{GRV} density function with respect to the scale
$\mu^{2}$, then to integrate over $k_{t}^{2}$ and compare the results with the
original GRV distribution. This was done because we will later use
equation (\ref{pdf2}) 
which relates the unintegrated parton density function
$\mathcal{G}(x,k_{t}^{2})$ to the integrated one $g(x,\mu^{2})$.
The adaptive Gaussian method taken from~\cite{GADAP} was used for the integration. As seen in
Figure~\ref{deriv2}, the DFRIDR routine reproduced the distribution
succesfully, except in the high $x$ region which will be discussed
later. However, with the same value of the parameter as used above, the DDERIV
approach gave unacceptable results, and only after changing this parameter again the obtained distribution (which is shown in the figure) started to resemble the original one. Because of this poor reproduction of the input distribution, and its sensitivity to the choice of parameter, we decided to use DFRIDR instead of DDERIV.

\begin{figure}
\begin{center}
\resizebox{12cm}{!}{\includegraphics{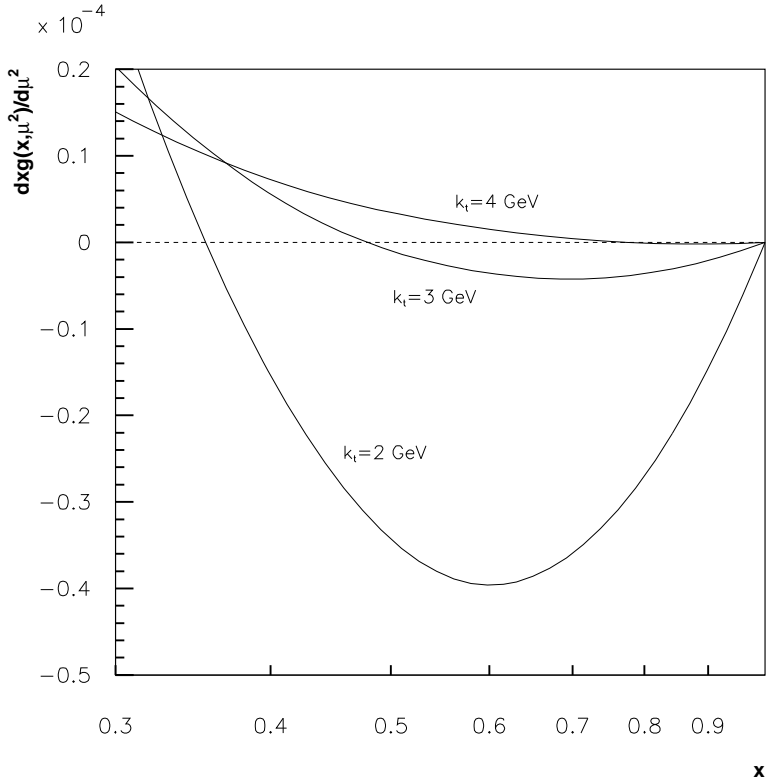}}
\caption{\it $\mathcal{G}(x,k_{t}^{2})$ as a function of $x$ for
different values of $\mu =k_{t}$. When the function becomes negative, the differentiation method of obtaining the unintegrated gluon density becomes invalid.}  
\label{deriv3}
\end{center}
\end{figure}

The deviations from the original distribution at high $x$ is due to
the fact that the function~(\ref{pdf2})
becomes negative in this region. This is because of the scaling
violation: as the scale $\mu ^{2}$ increases, smaller distances are
resolved and as a consequence we resolve more gluons with small $x$
(i.e. $x<0.1$)
and less gluons with high $x$. In Figure~\ref{deriv3} is shown how the
point, where the derivative becomes negative, changes with $\mu=k_{t}$. The smallest $x$-value for this point is obtained with
$k_{t}=2$~GeV, where $x\approx 0.35$.  Since $\mathcal{G}(x,k_{t}^{2})$ is interpreted as a probability, the method becomes invalid at such high $x$.

\subsection{Evolution of the unintegrated gluon density}
The evolution of the unintegrated gluon density was made using the
CCFM equation in a
forward evolution procedure based on the Monte Carlo technique as
described in~\cite{pdffit}. An initial gluon distribution is chosen
as well as a starting scale $\mu _{s}$ and a rescaled transverse momentum $\bar{q}_{s}$,
which is related to the emission angle (see equations (\ref{emang1})
and (\ref{emang2})), for the first emitted gluon. The default values
$\mu _{s}=\bar{q}_{s}=1.4$~GeV was obtained by a fit to the proton structure
function~\cite{pdffit}. The evolution then
starts with $\bar{q}_{max,0}=\bar{q}_{s}$, and $k_{t0}$ chosen from a Gaussian distribution around
$\mu _{s}$ to simulate a Fermi motion of the partons. $x_{0}$ is
chosen from the initial gluon distribution using the Monte Carlo
method. The branching is then repeated until the next emitted gluon
would have a higher $\bar{q}$  than $\bar{q}_{max,0}$. This is
repeated $10^{7}$ times for each of the 50 different maximum
$\bar{q}_{max,i}$ chosen from $\bar{q}_{max,0}=\bar{q}_{s} < \bar{q}_{max,i} <
\bar{q}_{max,50}=1800$~GeV in steps of $log(\bar{q}_{max})$, thus giving a
distribution $\mathcal{A}(x, k_{t}^{2}, \bar{q}^{2})$ in a $50\times
50\times 50$ grid of $log(x)$, $log(k_{t}^{2})$ and
$log(\bar{q}_{max})$. For values between the gridpoints the method of linear
interpolation is used. With the unintegrated parton distribution
avaliable, a backward evolution scheme can be used for the simulation
of the parton emissions. 

\subsection{One-loop evolution}

\begin{figure}
\begin{center}
\resizebox{14cm}{!}{\includegraphics{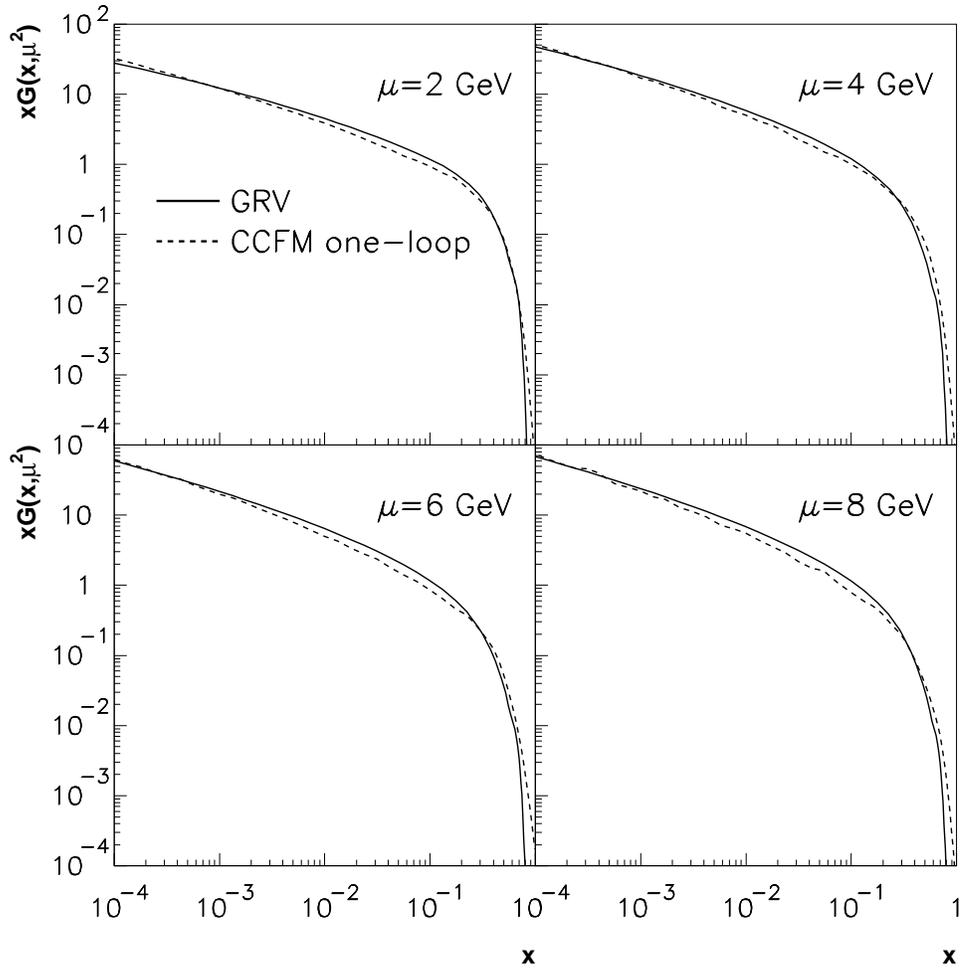}}
\caption{\it The integrated CCFM one-loop evolved distribution for the proton
compared to the GRV distribution at different scales.}
\label{protondglap}
\end{center}
\end{figure}

To make sure that the evolution procedure is working, the evolution
was first made for the proton with a GRV\footnote{GRV 98 LO was used}
input distribution (using the default values $\mu
_{s}=\bar{q}_{s}=1.4$~GeV) and using the one-loop
approximation~\cite{Webber, Marchesini, Gawron}, which means that the CCFM
evolution is reduced to the DGLAP evolution, with the difference that the
CCFM one-loop evolution still treats the full kinematics, i.e. it
takes the transverse momentum $k_{t}$ into account. This is done by
putting $\Delta _{ns}=1$ in the splitting function
(\ref{ccfmsplitting}) and reducing the condition (\ref{qcond})
to ordering in transverse momentum (\ref{DGLAPkt}). The obtained distribution $x\mathcal{A}(x, k_{t}
^{2},\bar{q}^{2})$ was then integrated over $k_{t}^{2}$,
\begin{equation}
\int _{0}^{\bar{q}^{2}}dk_{t}^{2} x\mathcal{A}(x, k_{t}^{2}, \bar{q}^{2}) = x\mathcal{G}(x, \bar{q}^{2}),
\end{equation}
and compared to the original GRV distribution at different scales, see Figure~\ref{protondglap}.
The small differences are due to the fact that the splitting function
only contains the singular parts, and that only the
splitting $g\rightarrow gg$ is considered. Nevertheless, the
distributions are still in good agreement which indicates that the
evolution machinery is working. 

\begin{figure}
\begin{center}
\resizebox{14cm}{!}{\includegraphics{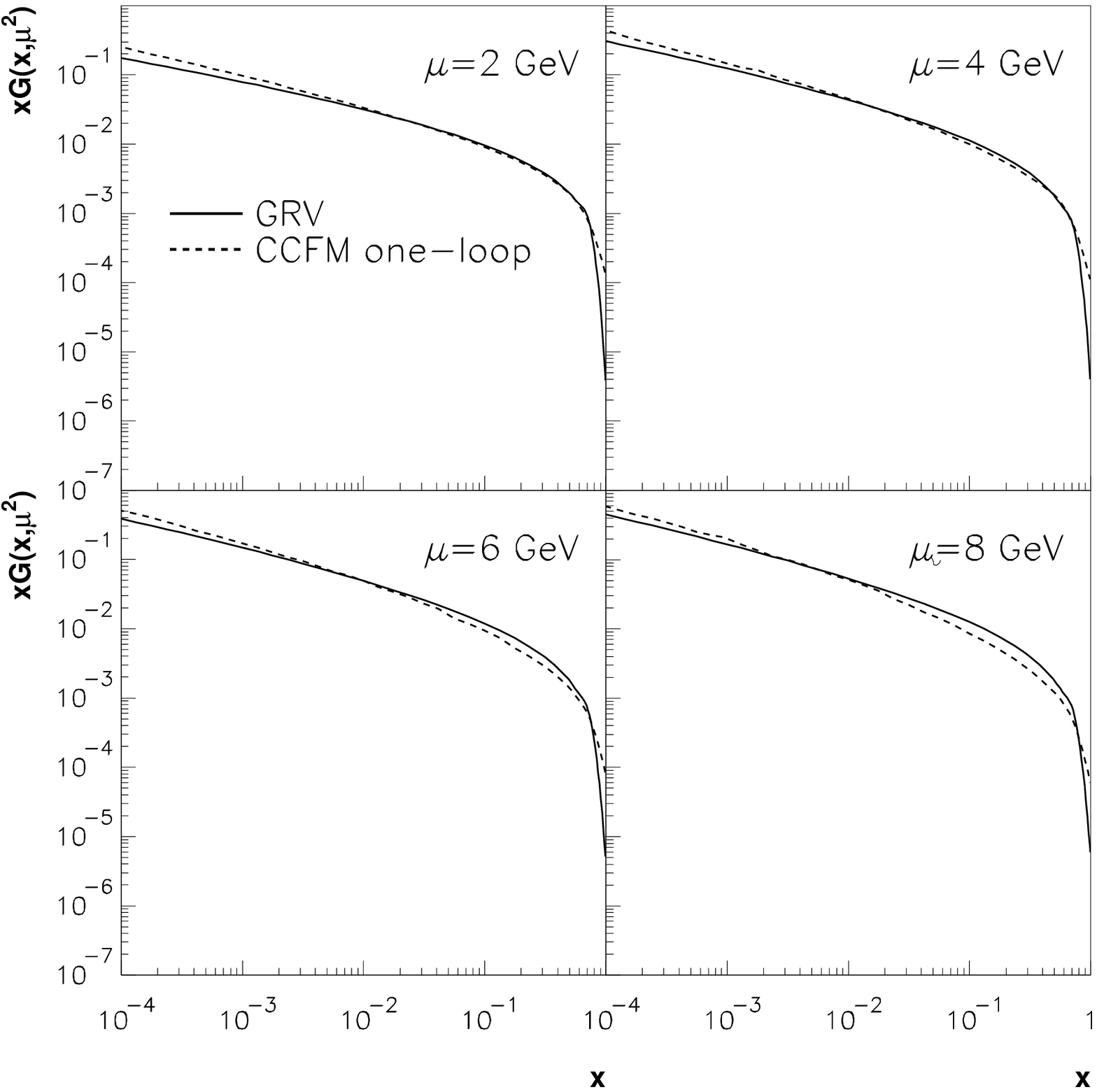}}
\caption{\it The integrated CCFM one-loop evolved distribution for the photon compared
to the GRV distribution at different scales.}
\label{photondglap}
\end{center}
\end{figure}

The situation for the photon is a bit different, since the structure
function, in addition to the hadronic component, now also consists of
a pointlike component which reflects the splitting of the photon into a
quark-antiquark pair. Also, there are no sum rules equivalent to those
in the proton case that constrain the quark distributions in the
photon. However, these differences do not matter, since we only use
the gluon distribution and the gluon splitting $g\rightarrow
gg$. Figure~\ref{photondglap} shows the evolved distribution for the
photon, using the same parameters as in the proton case and the
GRV\footnote{GRV-G LO was used} as input distribution, compared to
the input distribution at different scales. The one-loop evolved and
the GRV distributions
show a similar behaviour, although the one-loop
evolved distribution is a bit steeper at large scales. It is not
surprising that the one-loop evolution does not give equally good
agreement in the proton and the photon case, since the same parameters
have been used in both one-loop evolutions, while different techniques and
starting scales for the proton and the photon were used to obtain the
original GRV distributions. 

From the above we conclude that the one-loop evolution is able to
reproduce the input distributions reasonably well, and hence that the evolution machinery is
working for the CCFM one-loop case. A similar test cannot be done for
the full CCFM evolution, since the obtained distributions should be
different from the ones obtained with CCFM one-loop evolution. However, it has already
been shown \cite{pdffit} that the CCFM evolution is working for the
proton, in the sense that it can explain data, for example $b\bar{b}$
production at the TEVATRON, which could not be
explained with other methods. Since the only difference between the photon and proton
evolution is the starting distribution, it can be assumed that the
CCFM evolution will work also for the photon.

\subsection{CCFM evolution}
\begin{figure}
\begin{center}
\resizebox{14cm}{!}{\includegraphics{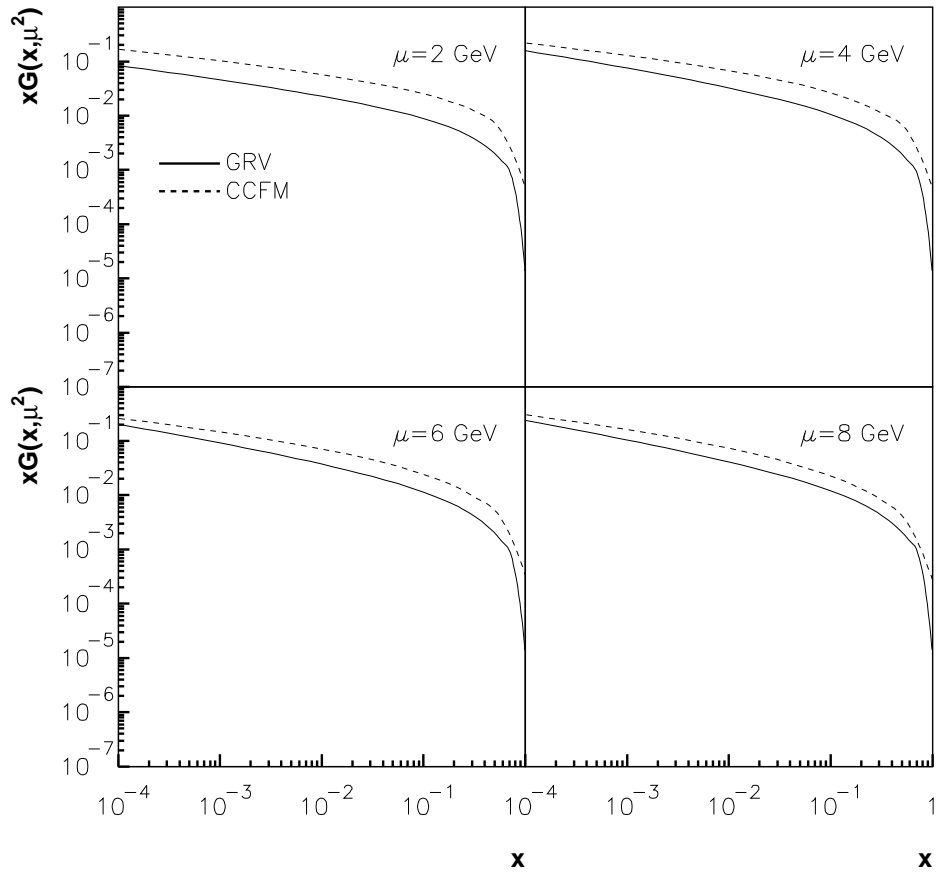}}
\caption{\it The integrated CCFM evolved distribution for the photon
compared to the GRV distribution at different scales.}
\label{photonccfm}
\end{center}
\end{figure}

After making sure that the parton evolution scheme gives consistent results, the unintegrated
gluon density for the photon was evolved using the full CCFM evolution
procedure. The obtained distributions integrated over $k_{t}^{2}$ are compared in Figure~\ref{photonccfm} to the
GRV\footnote{GRV-G HO was used} distribution, which was used as input
distribution. One of the more obvious differences is that the CCFM
evolved distribution is much higher than the original GRV. This is due
to differences in the definition of the integrated gluon density; one
can either integrate over $d^{2}k_{t}$ or over $dk^{2}_{t}$. The
difference is a factor $\pi$, which explains the main part of the
normalization difference. We can also note that the CCFM evolved
distribution is less steep than the original GRV. This is a
consequence of that the non-Sudakov form factor $\Delta
_{ns}\rightarrow 0$ when $z\rightarrow 0$ and thereby screens the $\frac{1}{z}$ factor in the splitting function. This results in a lower splitting probability compared to the DGLAP case (where $\Delta _{ns}=1$) which becomes visible in the small $x$ (small $z$) region where the factor $\frac{1}{z}$ dominates the splitting function.

\begin{figure}
\begin{center}
\resizebox{12cm}{!}{\includegraphics{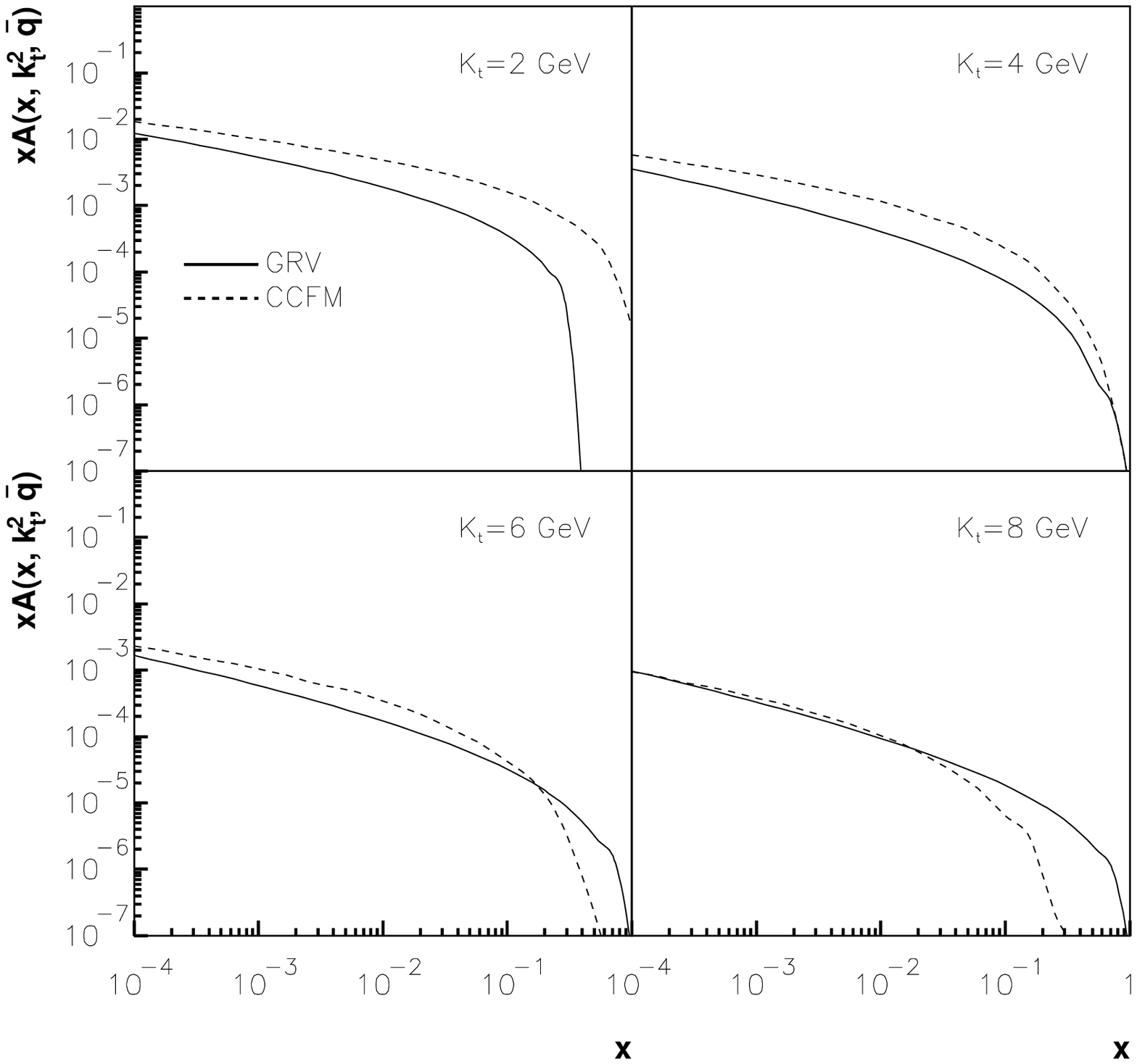}}
\resizebox{12cm}{!}{\includegraphics{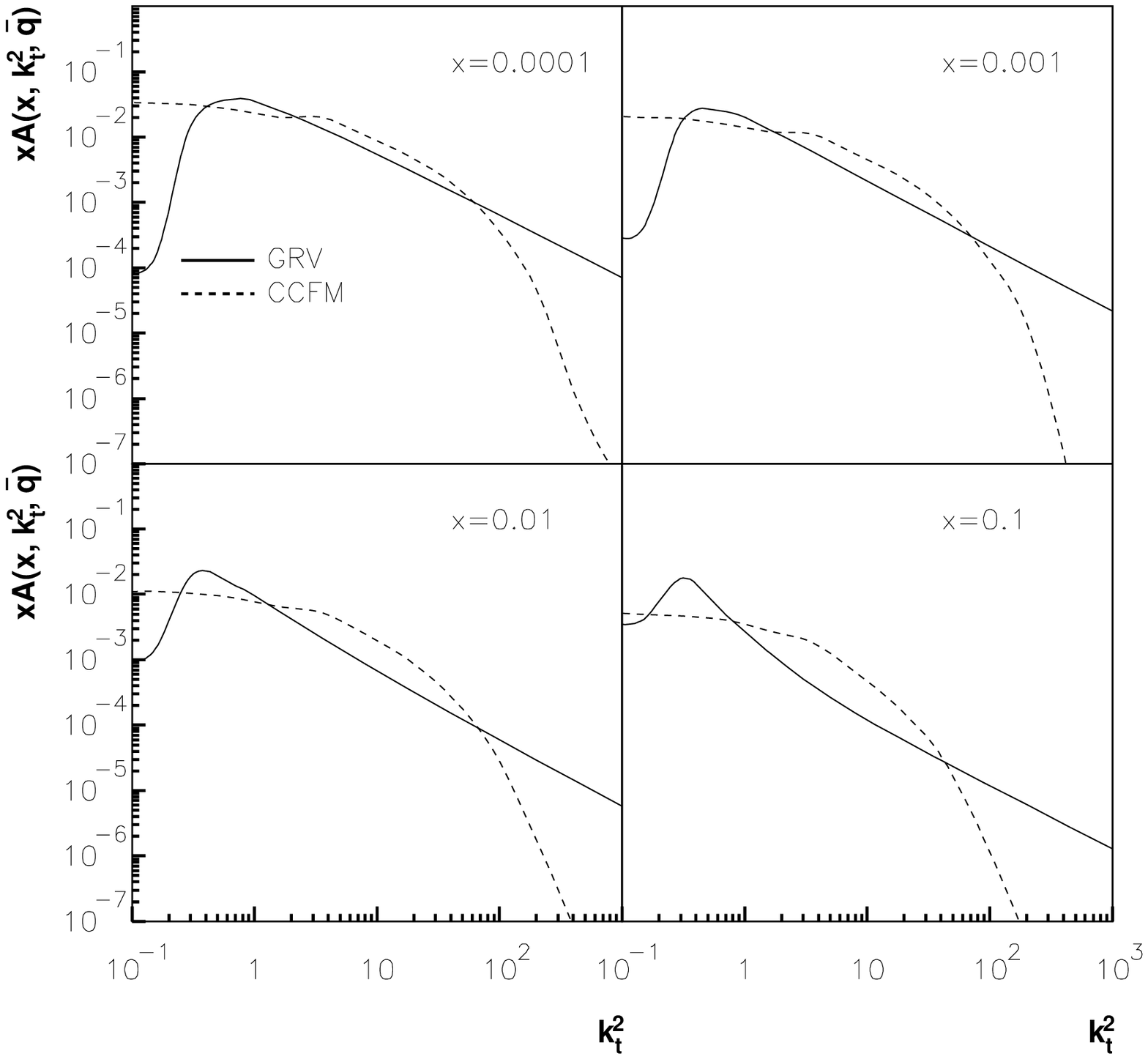}}
\caption{\it The unintegrated CCFM evolved distribution for the photon
compared to the derivative of the GRV distribution as a function of
$x$ for different values of $k_{t}$ (a), and as a function of
$k_{t}^{2}$ for different values of $x$ (b).}
\label{photonccfmun}
\end{center}
\end{figure}

In Figure~\ref{photonccfmun} is shown the unintegrated
distribution as a function of $x$ for different values of $k_{t}$, and
as a function of $k_{t}^{2}$ for different values of $x$, compared to the
derivative of GRV. Here,
$\mu =\bar{q}=10$~GeV. Also in Figure~\ref{photonccfmun}a we see that the
CCFM distribution in general is higher (except at very high $k_{t}$)
and we can also see the non-Sudakov supression at small $x$. We should
also remember that the derivative method becomes invalid at high $x$
because $\frac{dg(x, \mu ^{2})}{d\mu ^{2}}$ becomes negative. This
explains the big difference between the CCFM evolved distribution and the GRV at $k_{t}=2$ GeV, because the derivative becomes negative already at $x \approx 0.35$ (see Figure~\ref{deriv3}).
In Figure \ref{photonccfmun}b one can see the typical exponential
$k_{t}$-dependence of the derivative of GRV. Since the GRV
distribution is only defined for $\mu ^{2}=k_{t}^{2}\geq 0.3$ GeV$^{2}$,
it is replaced with $\frac{x\mathcal{G}(x, k_{t0}^{2})}{k_{t0}^{2}}$
for $k_{t}^{2}<k_{t0}^{2}=0.3$ GeV$^{2}$ \cite{Baranov}. We can also
note that the probability for the gluons to have a
$k_{t}^{2}>\mu ^{2}=100$ GeV$^{2}$ in the CCFM approach is significant, which would
not be possible using DGLAP. This is a consequence of the $k_{t}^{2}$-ordering in the DGLAP
evolution, where $k_{t}^{2}$ is always less than $\mu ^{2}$, see
(\ref{DGLAPkt}), and the non-ordering in $k_{t}^{2}$ in the CCFM
evolution, where the gluons can have any kinematically allowed value of $k_{t}^{2}$. 

\subsection{Uncertainties}
\label{unc1}
\begin{figure}
\begin{center}
\resizebox{14cm}{!}{\includegraphics{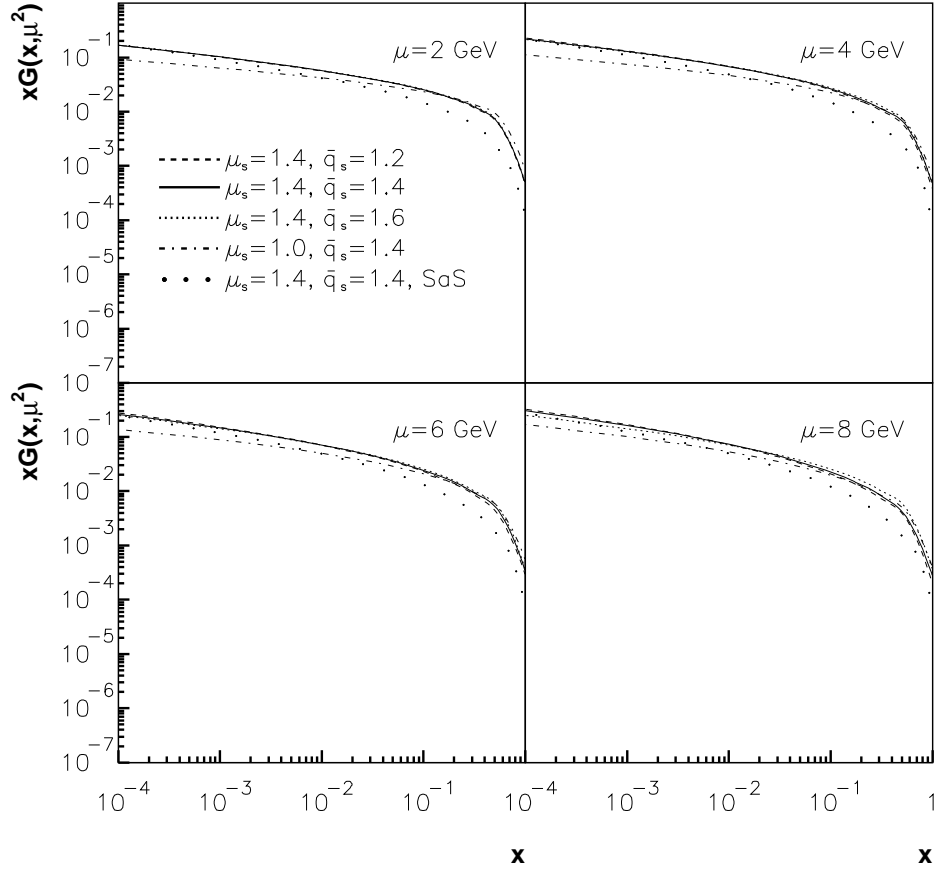}}
\caption{\it The integrated CCFM evolved distributions for the photon with
different combinations of $\mu_{s}$, $\bar{q}_{s}$ and input distributions.}
\label{uncertain}
\end{center}
\end{figure}

\begin{figure}
\begin{center}
\resizebox{12cm}{!}{\includegraphics{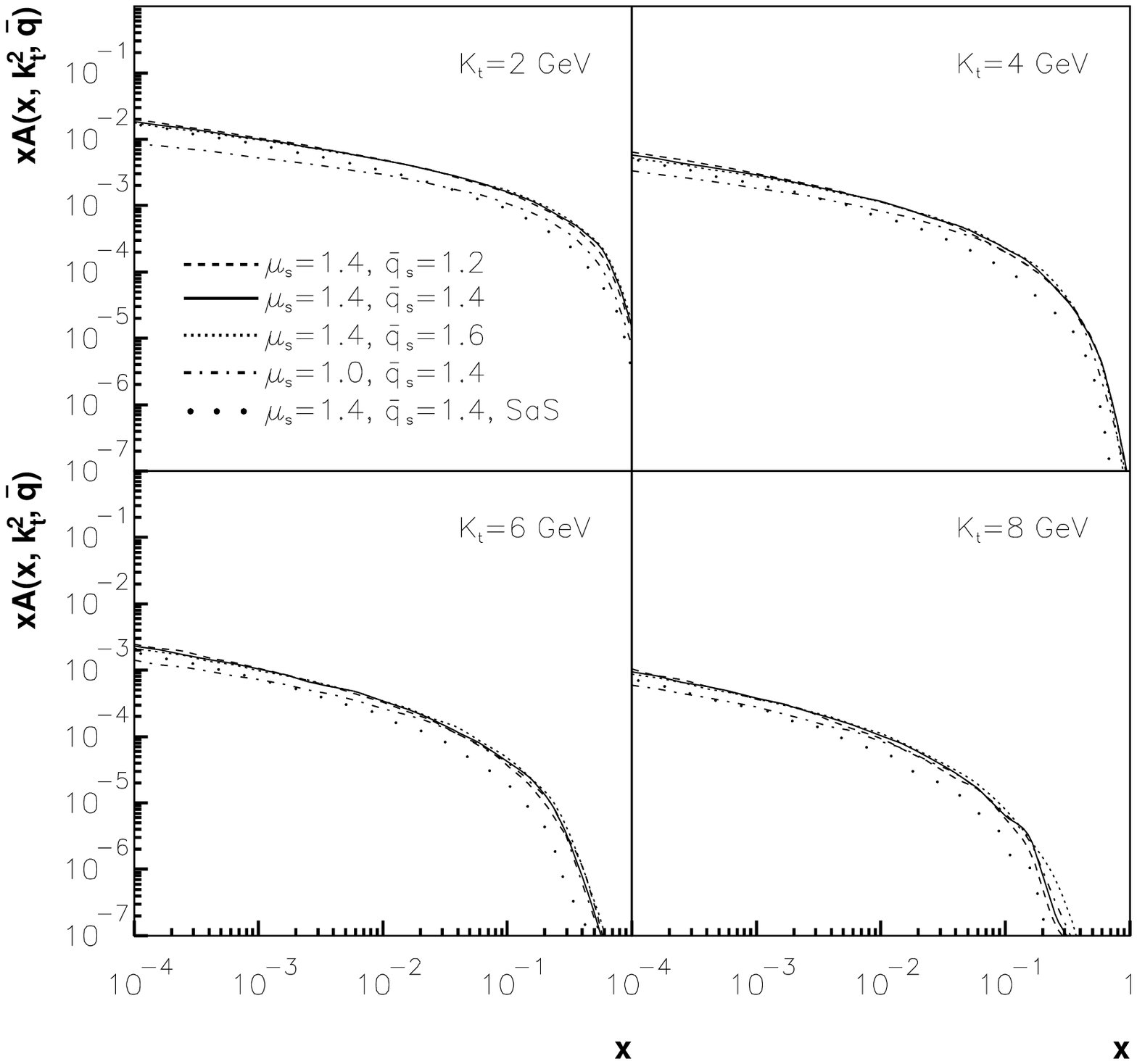}}
\resizebox{12cm}{!}{\includegraphics{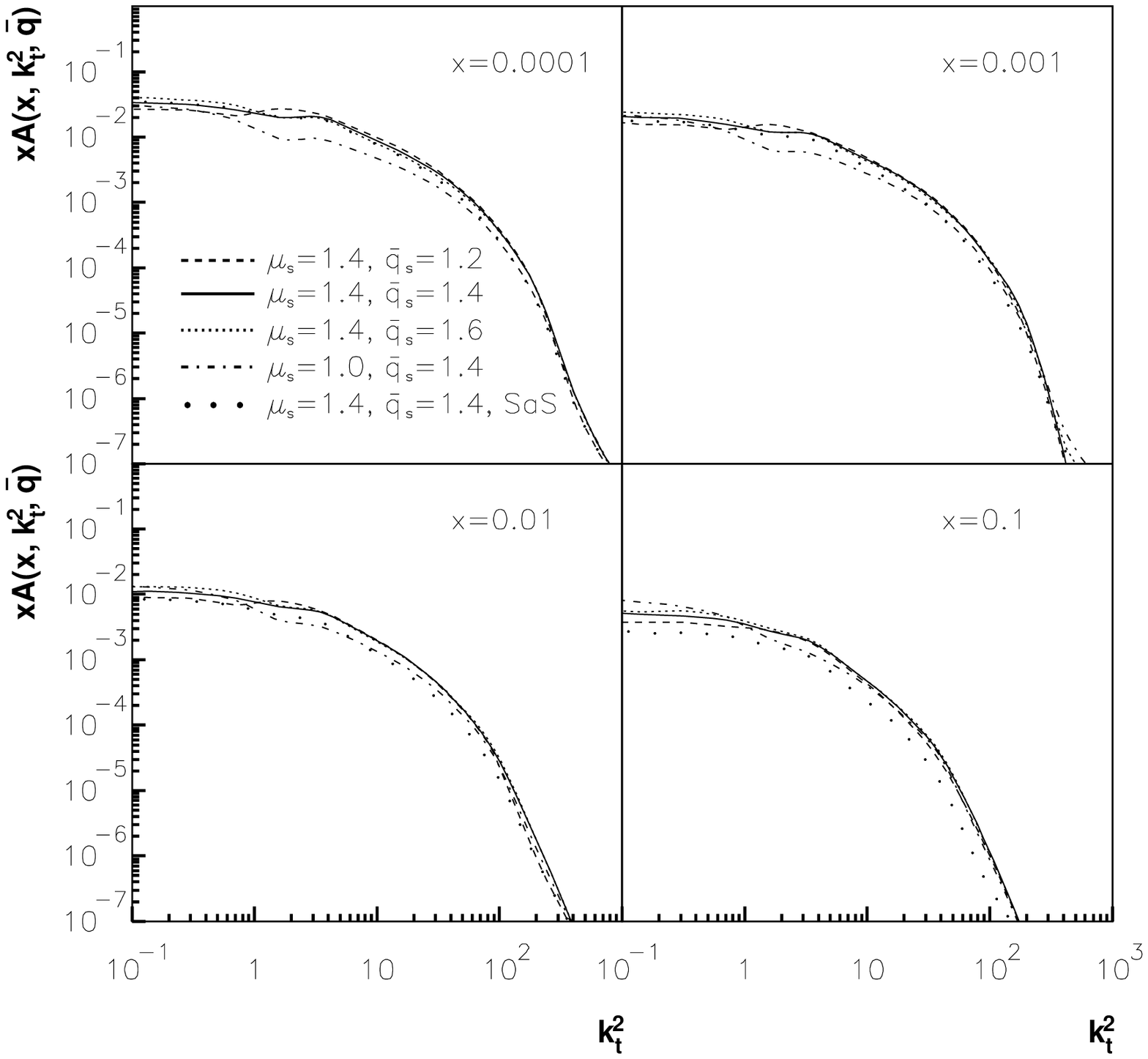}}
\caption{\it The unintegrated CCFM evolved distributions for the photon with
different combinations of $\mu_{s}$, $\bar{q}_{s}$ and input distributions. }
\label{uncertain_un}
\end{center}
\end{figure}

\begin{figure}
\begin{center}
\resizebox{14cm}{!}{\includegraphics{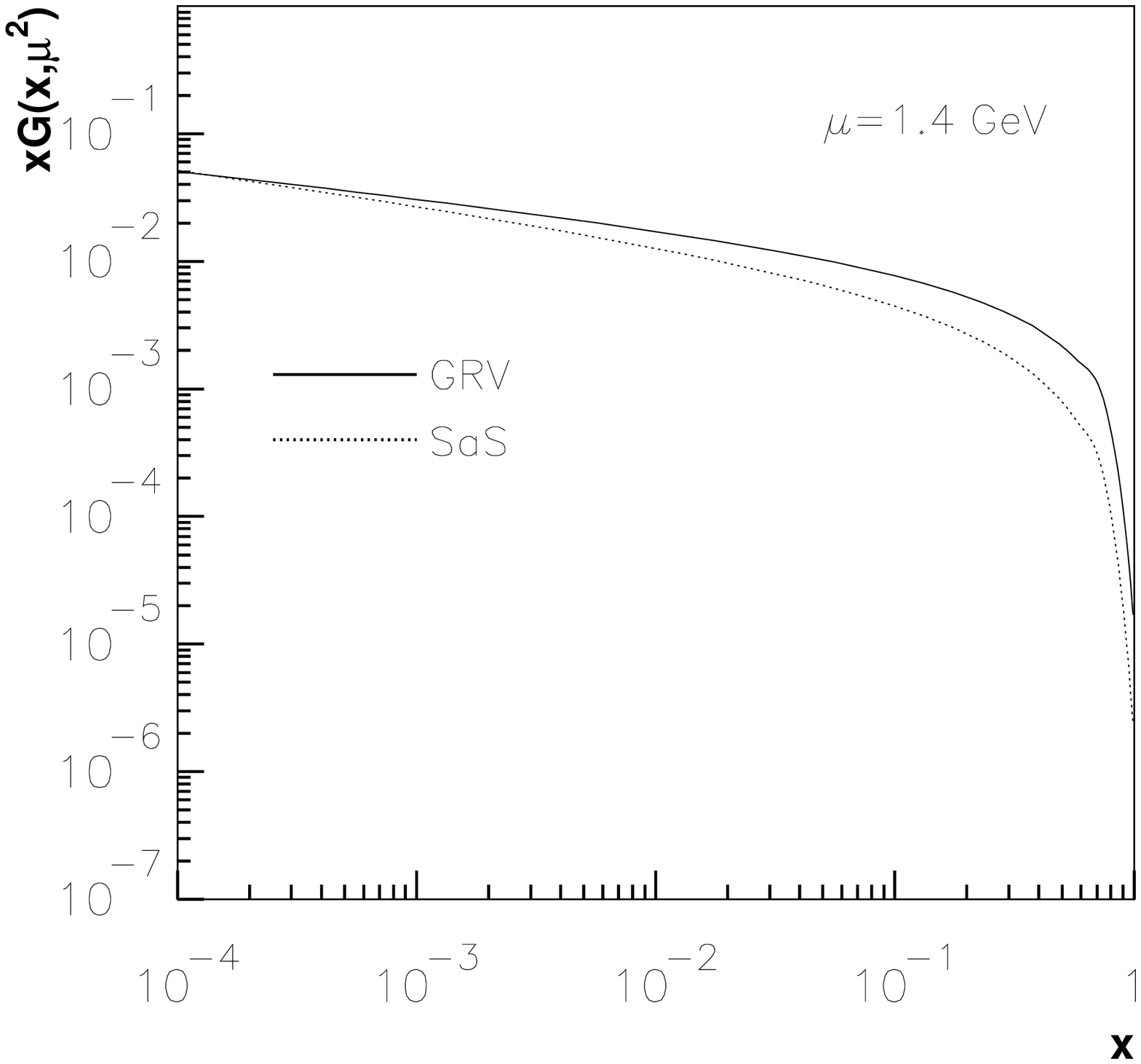}}
\caption{\it The GRV and SaS distributions compared at the starting scale $\mu
=1.4$~GeV.}
\label{originalpdf}
\end{center}
\end{figure}

In the evolution of the unintegrated gluon distributions above, the
GRV density was used as input distribution, with $\mu _{s}=\bar{q}_{s}=1.4$~GeV given from a fit to the parton
density of the proton \cite{pdffit}. A natural question is how
sensitive the results are to the selected starting distribution and
the parameters chosen. Therefore,
the evolution was made also with the SaS\footnote{SaS-G 1D was used}
\cite{SaS} parton distribution as input (and keeping $\mu
_{s}=\bar{q}_{s}=1.4$~GeV), and also with the GRV parametrization as input but with
$\mu _{s}=1.4$~GeV and $\bar{q}_{s}=1.2$~GeV, $\mu _{s}=1.4$~GeV and
$\bar{q}_{s}=1.6$~GeV, and $\mu _{s}=1$~GeV and
$\bar{q}_{s}=1.4$~GeV. The various distributions obtained are compared
to the previous ones in Figure~\ref{uncertain} and~\ref{uncertain_un}.
When comparing the integrated distributions (Figure~\ref{uncertain}) we see that there is only a weak dependence on
the starting value $\bar{q}_{s}$, such that the distribution becomes
somewhat flatter at large $\bar{q}_{s}$. This can
be understood, since a larger starting angle would mean that it takes
less emissions to reach the maximum value (since there is angular
ordering), and less emissions would mean that the gluon entering the
matrix element (gluon $n$ in Figure~\ref{ladder}) would have a larger
longitudinal momentum fraction $x$. A change in $\mu _{s}$
makes a significant difference, where a smaller $\mu _{s}$ gives a
distribution which is lower at small $x$. One can also see that the
evolution depends on the choice of input distribution, where the main
features of the input distribution remains after the evolution, see also Figure~\ref{originalpdf}.

Also when comparing the unintegrated gluon densities in
Figure~\ref{uncertain_un}, we see that the dependence on $\bar{q}_{s}$
is small while changing $\mu _{s}$ makes a significant difference,
especially in the region $1<k_{t}^{2}<10$. This is probably due to
details in the evolution: In order to avoid soft emissions a cut-off
scale $\mu _{s}$ is defined. Thus if the propagator gluon has
$k_{t}>\mu _{s}$ a real emission is allowed, but it is not if
$k_{t}<\mu _{s}$. Since there is no ordering in $k_{t}$ in the CCFM
evolution any propagator gluon along the evolution chain might get
$k_{t}<\mu _{s}$ after a real emission. In that case, the evolution of
the propagator is continued until $k_{t}>\mu _{s}$ and a real emission
is allowed.


\section{Heavy quark cross sections}
\label{mycross}
The obtained unintegrated gluon density was used as an input to the MC generator CASCADE
to calculate heavy quark cross sections in $e^{+}e^{-}$ collisions,
which are compared to LEP data. Also, predictions for heavy
quark cross sections in $e^{+}e^{-}$ and $\gamma\gamma$ collisions at
TESLA energies are given.

\subsection{Cross section calculation test}
The $e^{+}e^{-}$ cross sections can be calculated using the relation
\begin{equation}
\sigma _{e^{+}e^{-}\rightarrow e^{+}e^{-}b\bar{b}, c\bar{c}X}=\int
\frac{dQ_{1}^{2}}{Q_{1}^{2}}\frac{dQ_{2}^{2}}{Q_{2}^{2}}\mathcal{L}_{\gamma
\gamma}(W_{\gamma \gamma}, Q_{1}^{2}, Q_{2}^{2})\sigma _{\gamma \gamma
\rightarrow b\bar{b}, c\bar{c}X}(W_{\gamma \gamma})dW_{\gamma \gamma}
\end{equation}
where $\mathcal{L}_{\gamma \gamma}$ is the photon flux in the electron, $W_{\gamma
\gamma}$ is the center of mass energy in the $\gamma \gamma$ system,
$Q_{i}$ is the virtuality of photon $i$ and $\sigma _{\gamma \gamma
\rightarrow b\bar{b}, c\bar{c}X}$ is the cross section for the
subprocess $\gamma \gamma \rightarrow b\bar{b}, c\bar{c}X$. 

The direct and single resolved cross
sections for $b$ and $c$ quark production in $e^{+}e^{-}$ collisions were calculated with CASCADE
and compared with the results in \cite{Drees}. Since the direct cross
section is independent of the gluon content in the photon, and hence
independent of evolution equations, these cross sections should agree. In heavy flavour production, CASCADE uses off-shell matrix elements as
calculated in~\cite{catani}. Errors were found in~\cite{catani} (eq. B16) where a factor
$N_{c}e_{q}^{2}$ was missing in the direct cross section ($N_{c}=3$
being the color factor and $e_{q}$ being the EM charge of the produced
quarks). With these factors taken into account, the direct cross
sections from CASCADE and from leading and next-to-leading order calculations
\cite{Drees} were in good agreement, see Table~\ref{crosstest}. For the single
resolved case, the DGLAP evolution was used with the gluon density according to GRV\footnote{GRV-G
1HO was used, as in \cite{Drees}} to describe the photon structure and
with cuts and parameters as in \cite{Drees}. As seen, also these cross sections
where in good agreement. The aim of this comparison is just to show
that the basic machinery for cross
section calculations in CASCADE is working.

\begin{table}[h]
\begin{center}
\begin{tabular}{ccccc}
 & \multicolumn{4}{c}{\emph{CHARM}} \\
$\sqrt{s}$ & \multicolumn{2}{c}{\emph{$\sigma _{DIR} (pb)$}} & \multicolumn{2}{c}{\emph{$\sigma _{1-RES} (pb)$}} \\
\emph{(GeV)} & \emph{DREES} & \emph{CASCADE} & \emph{DREES} & \emph{CASCADE}\\ \hline
180 & 296.8 (396.3) & 328 & 269.9 (305.8) & 217\\
91.2 & 191.2 (254.8) & 213 & 93.38 (109.5) & 80\\
\end{tabular}
\begin{tabular}{ccccc}
 & \multicolumn{4}{c}{} \\
 & \multicolumn{4}{c}{\emph{BEAUTY}} \\
$\sqrt{s}$ & \multicolumn{2}{c}{\emph{$\sigma _{DIR} (pb)$}} & \multicolumn{2}{c}{\emph{$\sigma _{1-RES} (pb)$}} \\
\emph{(GeV)} & \emph{DREES} & \emph{CASCADE} & \emph{DREES} &
\emph{CASCADE}\\ \hline
180 & 1.280 (1.582) & 1.35 & 1.371 (1.644) & 1.19\\
91.2 & 0.663 (0.823) & 0.71 & 0.302 (0.371) & 0.27 \\
\end{tabular}
\caption{\it  Direct and single resolved charm and beauty cross
sections
calculated with CASCADE and compared to leading and next-to-leading
order (in parenthesis) calculations in \cite{Drees} (DREES). Here,
$\mu ^{2}=2m^{2}$, $\Lambda =0.34$~GeV, $m_{b}=4.75$~GeV and
$m_{c}=1.6$~GeV. In CASCADE, the scale factor 0.5 was used to obtain
the correct scale.}
\label{crosstest}
\end{center}
\end{table}

\subsection{$e^{+}e^{-}$ cross sections at LEP}
\begin{figure}
\begin{center}
\resizebox{14cm}{!}{\includegraphics{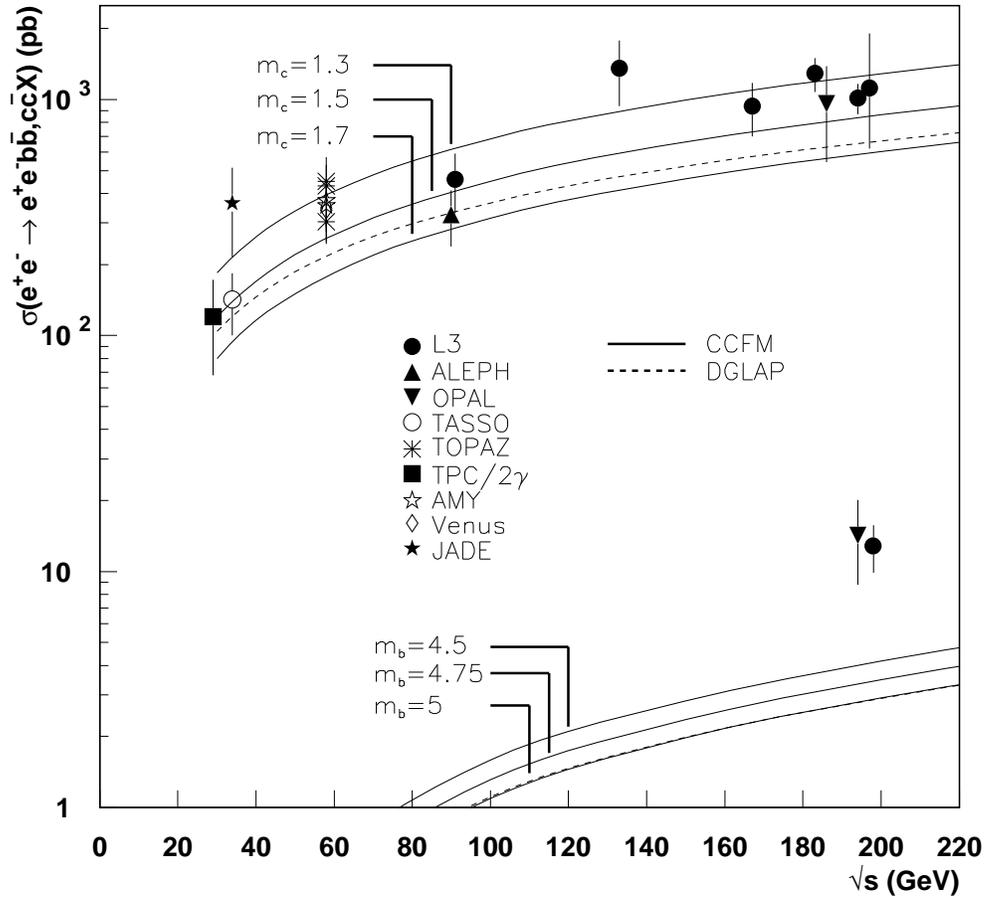}}
\caption{\it The charm and beauty cross sections calculated with
$k_{t}$-factorization compared to the DGLAP approach and data.}
\label{eexsec}
\end{center}
\end{figure}

In Figure~\ref{eexsec} are shown the obtained cross sections for $b$ and
$c$ quark production in $e^{+}e^{-}$ collisions calculated with the $k_{t}$-factorization approach, using off-shell ME and a CCFM unintegrated
gluon density evolved from a starting distribution given by GRV,
compared to the collinear (DGLAP)
predictions and data \cite{LEP1, LEP2, LEP7, LEP8, LEP9, LEP10}. Here, the masses $m_{b}=4.5, 4.75, 5$~GeV
and $m_{c}=1.3, 1.5, 1.7$~GeV were used for the CCFM cross sections, while
$m_{b}=4.75$~GeV and $m_{c}=1.5$~GeV were used for the cross sections
calculated with DGLAP. This variation of the quark masses gives an
estimate of the uncertainty in the calculation.
One can see that the $k_{t}$-factorization approach describes the
charm data reasonably well and that it gives larger cross sections
compared to the collinear method. One can also see that the choice of
mass makes a significant difference, such that a smaller mass gives a
larger cross section. Nevertheless, the cross sections obtained with
$1.3$~GeV$<m_{c}<1.7$~GeV all lie within the experimental
uncertainties. 

\begin{figure}
\begin{center}
\resizebox{14cm}{!}{\includegraphics{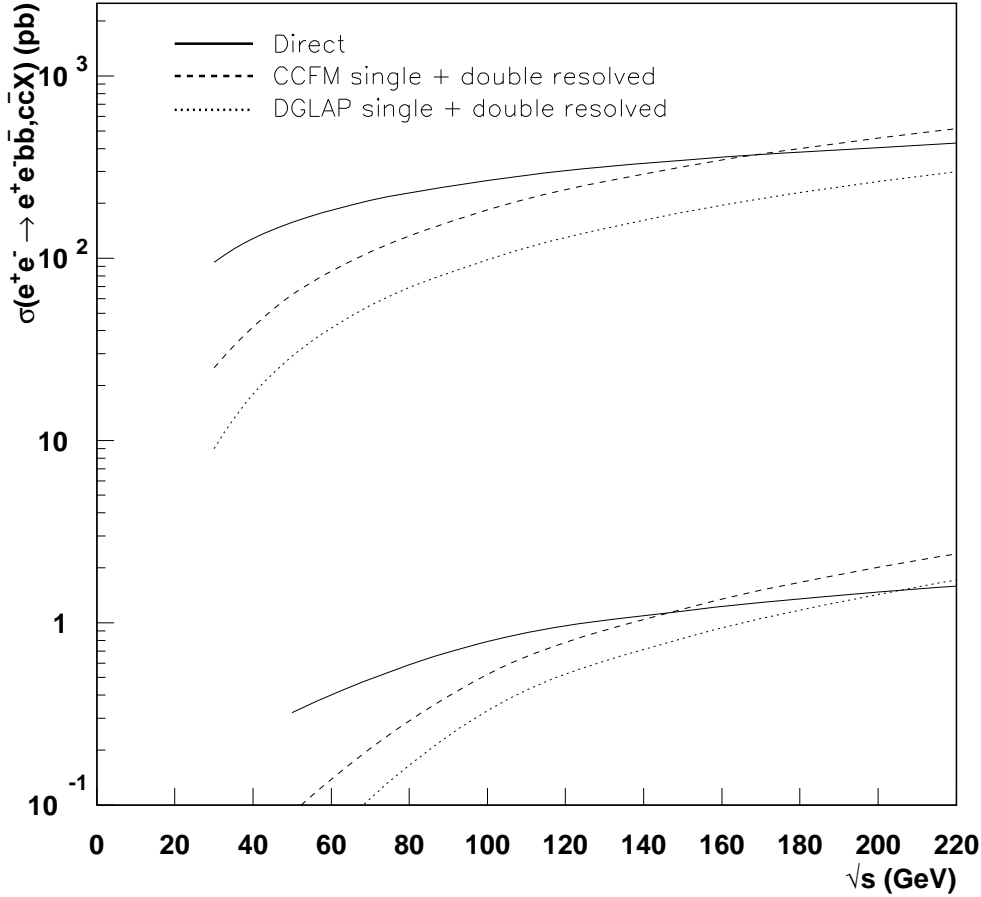}}
\caption{\it The direct and resolved
cross sections in heavy quark production calculated with CCFM and DGLAP.}
\label{xsecdirres}
\end{center}
\end{figure}

\begin{figure}
\begin{center}
\resizebox{14cm}{!}{\includegraphics{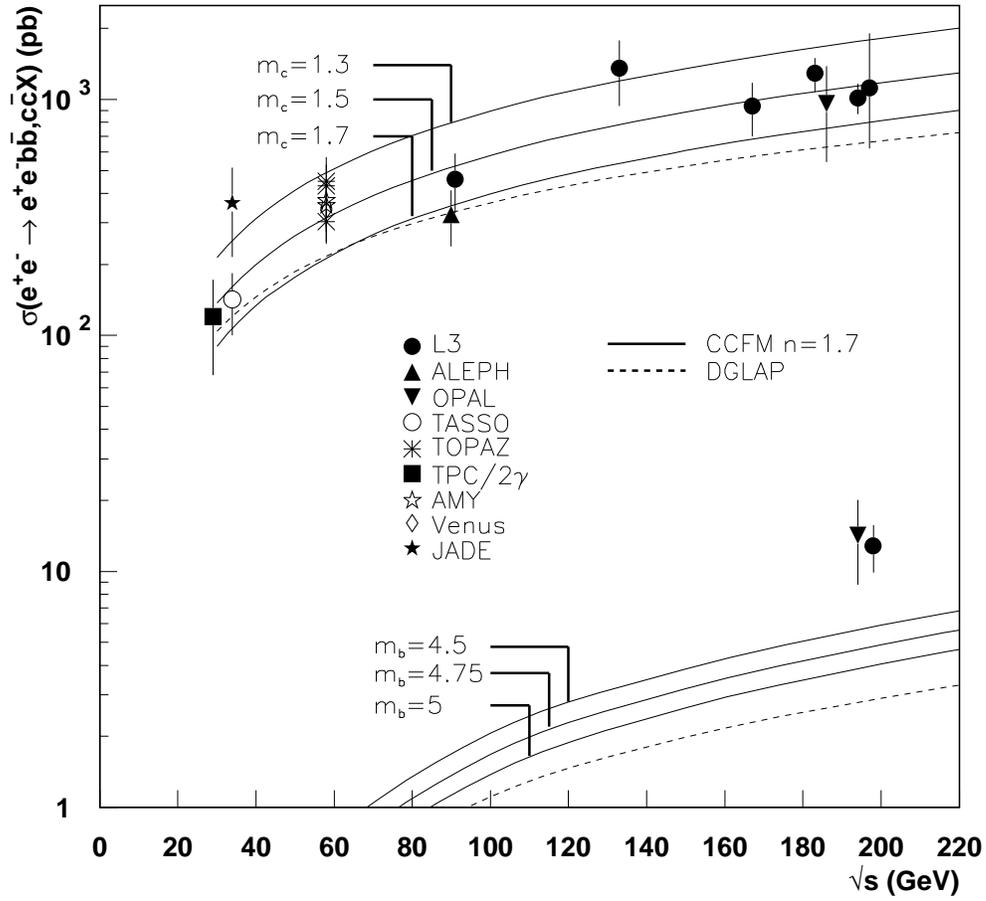}}
\caption{\it The cross sections for heavy quark production calculated
with CCFM and normalized to charm data at $\sqrt{s}=200$~GeV with a factor 
$n=1.7$, compared to DGLAP predictions and data.}
\label{xsecnorm}
\end{center}
\end{figure}

The cross sections for $b\bar{b}$ production, on the
other hand, are not well described. However, in the CCFM approach,
one has to determine carefully the normalization of the cross sections. The
reason for this is that the normalization of the unintegrated gluon
density does not have to be the same as for the standard gluon
densities generated according to the collinear scheme, since the
integration is made over different regions (i.e. the integral over
$dk_{t}^{2}$ is from $0$ to $\mu ^{2}$ in the collinear approach,
whereas in the CCFM formalism the integral is over all kinematically allowed $k_{t}^{2}$) and
the matrix elements are different. This normalization factor has to be
applied to the single and double resolved cross sections, but not
to the direct cross section since it is independent of the gluon density. The
contribution of the direct and resolved cross
sections are shown in Figure~\ref{xsecdirres}. As
seen, the differences between CCFM and DGLAP are due to differences in
the resolved contributions of the cross sections.
Normalizing the CCFM cross section to the charm data at
$\sqrt{s}=200$~GeV for $m_{c}=1.5$~GeV, by applying a normalization factor $n=1.7$ to the
single and double resolved cross sections, gives a much steeper
behaviour of the total cross section, as seen in Figure~\ref{xsecnorm}.
However, the CCFM predictions are still in good agreement with the charm data at $\sqrt{s}<200$~GeV. 
The normalization also improves the situation for the $b\bar{b}$ cross
section, where our calculations give $\sigma _{b\bar{b}}=4.9$~pb at
$\sqrt{s}=200$~GeV. This can be compared to standard collinear NLO
predictions which give $\sigma _{b\bar{b}}\approx 4$~pb \cite{LEP7} and
results based on~\cite{Motyka}, giving
$\sigma _{b\bar{b}}=1.9$~pb using a KMR gluon density, and $\sigma _{b\bar{b}}=2.7$~pb using a GBW gluon density at the same energy. The CCFM
cross sections are of the same order but a bit larger than NLO standard
calculations, which has also been observed in $ep$ collisions at
HERA. Thus, the improvement is
not sufficient fully to describe the $b\bar{b}$ data, which is still a
factor 2-3 above predictions.

\subsection{$e^{+}e^{-}$ and $\gamma \gamma$ cross sections at TESLA}
\begin{figure}
\begin{center}
\resizebox{14cm}{!}{\includegraphics{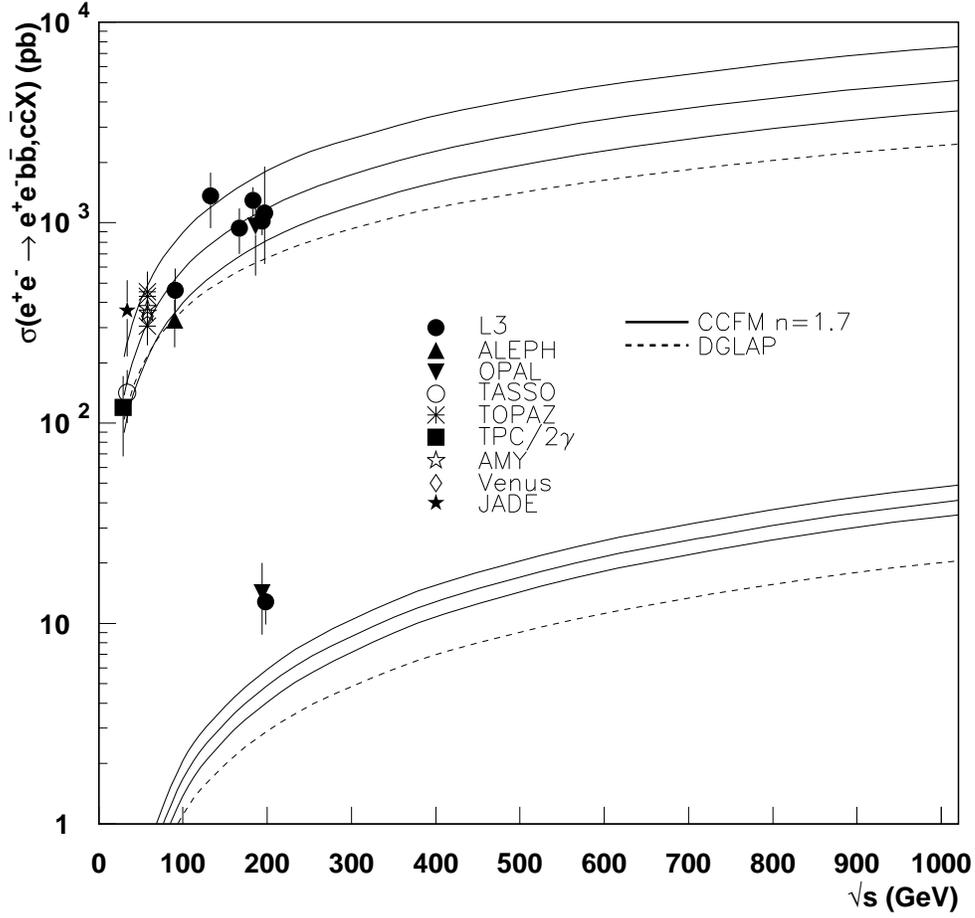}}
\caption{\it Heavy quark cross section predictions for $e^{+}e^{-}$
collisions at TESLA energies calculated with CCFM and compared to the
DGLAP predictions. For CCFM, the quark masses $m_{c}=1.3, 1.5, 1.7$~GeV
and $m_{b}=4.5, 4.75, 5$~GeV were used, where the smallest mass gives
the largest cross section. For DGLAP, $m_{c}=1.5$~GeV and $m_{b}=4.75$~GeV. Also, the normalization factor $n=1.7$ was
applied to the resolved cross sections in the CCFM calculations.}
\label{xsec2norm}
\end{center}
\end{figure}

The CCFM predictions for heavy quark production in $e^{+}e^{-}$
collisions at TESLA energies are shown in
Figure~\ref{xsec2norm}. Here, the normalization factor $n=1.7$ has
been applied for the resolved cross sections. With a charm mass of
$m_{c}=1.5$~GeV the CCFM predictions at $\sqrt{s}=500 \ (800)$~GeV are
\begin{center}
$\sigma _{e^{+}e^{-}\rightarrow e^{+}e^{-}c\bar{c}X}=2770 \ (4165)$~pb.
\end{center}
For beauty, the predictions are
\begin{center}
$\sigma _{e^{+}e^{-}\rightarrow e^{+}e^{-}b\bar{b}X}=17.06 \ (30.72)$~pb
\end{center}
using a beauty mass of $m_{b}=4.75$~GeV.

\begin{figure}
\begin{center}
\resizebox{14cm}{!}{\includegraphics{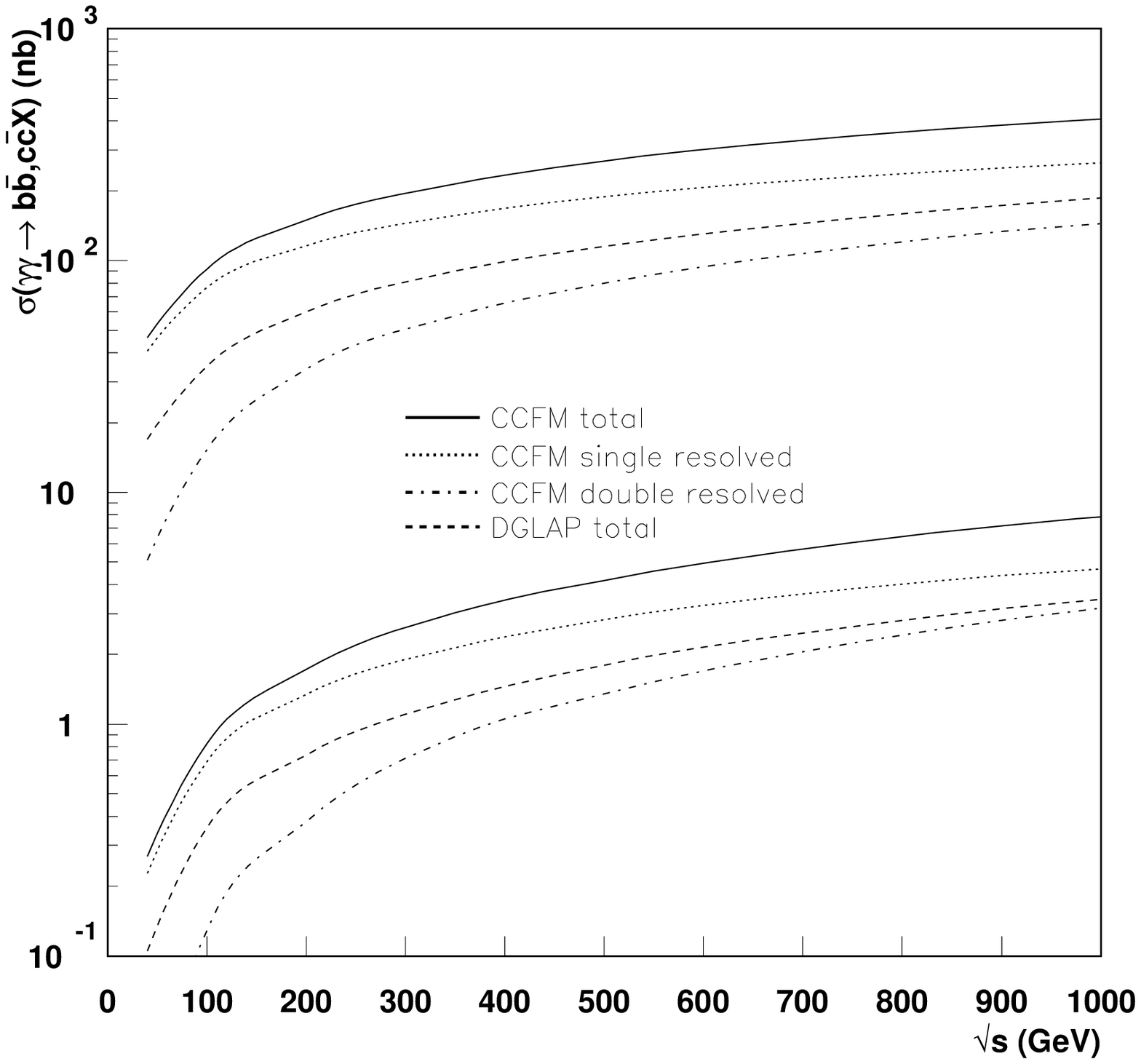}}
\caption{\it Heavy quark cross section predictions for $\gamma \gamma$
collisions at TESLA energies. The masses $m_{c}=1.5$~GeV and
$m_{b}=4.75$~GeV were used.}
\label{xsecgam}
\end{center}
\end{figure}

Figure~\ref{xsecgam} shows the charm and beauty cross sections in
$\gamma \gamma$ collisions at TESLA energies calculated using the CCFM
evolution. The normalization factor $n=1.7$ has been applied to the
resolved contributions. At $\sqrt{s}=500 \ (800)$~GeV, the predicted cross sections are
\begin{center}
$\sigma _{\gamma \gamma \rightarrow c\bar{c}X}=269 \ (359)$~nb
\end{center}
for charm and
\begin{center}
$\sigma _{\gamma \gamma \rightarrow b\bar{b}X}=4.18 \ (6.38)$~nb
\end{center}
for beauty. Also here, the masses $m_{c}=1.5$~GeV and $m_{b}=4.75$~GeV
were used. Figure~\ref{xsecgam} also shows the resolved contributions
of the cross sections. As seen, the single and double resolved cross
sections dominate the total cross section at these energies. The
reason for this is that the direct contribution, in contrast to the resolved
ones, decreases with energy, and is only about
1$\%$ of the total cross section at $\sqrt{s}=100$~GeV.

\subsection{Uncertainties}
\label{chapt_uncert}
\begin{figure}
\begin{center}
\resizebox{14cm}{!}{\includegraphics{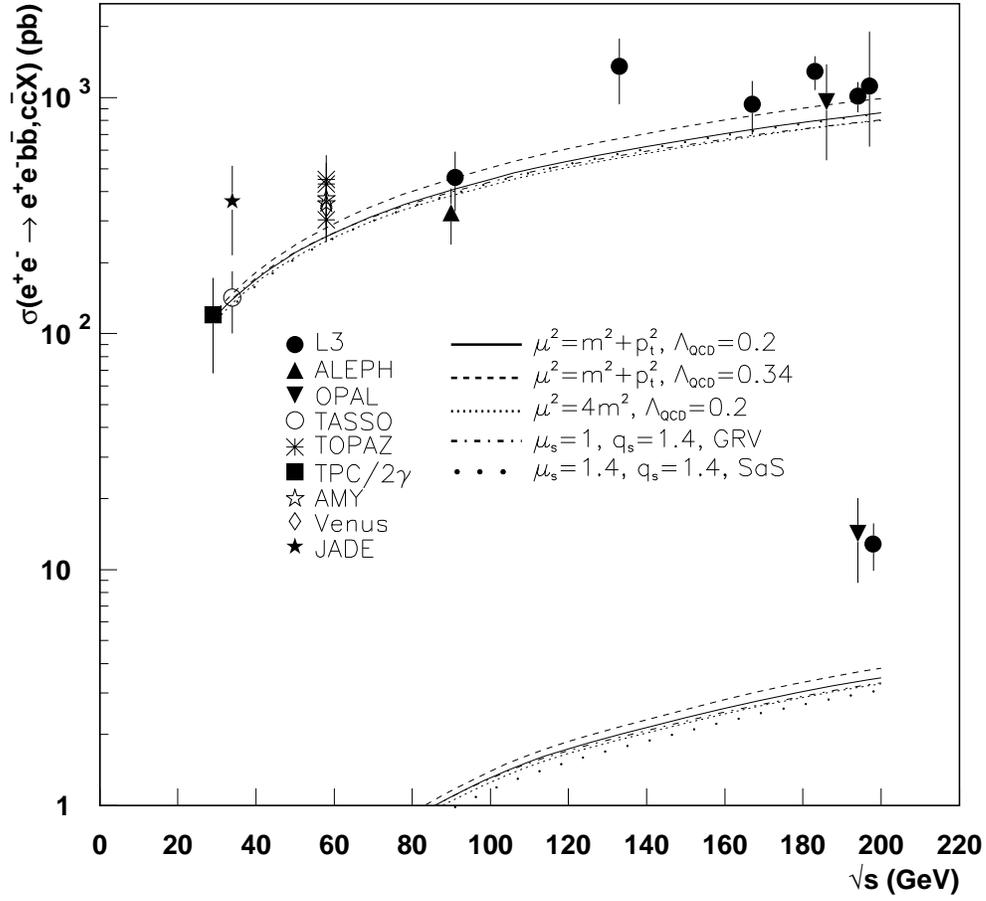}}
\caption{\it The cross section dependence on $\Lambda$, $\mu^{2}$ and
different unintegrated gluon densities. }
\label{xsecvar}
\end{center}
\end{figure}

The cross section for heavy quark production depends on the
unintegrated gluon density and the partonic cross section. These in turn depend on a
few parameters which are not fixed by theory but have to be determined
from fits to experimental data, and
therefore give rise to some uncertainties in the calculations. These
uncertainties are discussed below and shown in Figure~\ref{xsecvar}
for $e^{+}e^{-}$ collisions.

\textbf{The CCFM evolution} of the unintegrated gluon density is sensitive to the input
gluon density, the starting scale $\mu _{s}$ and the starting
value for the rescaled transverse momentum $\bar{q}_{s}$. The default values in this analysis 
are $\mu _{s}=\bar{q}_{s}=1.4$~GeV, obtained
from a fit to the parton distribution in the proton \cite{pdffit}, with a
GRV input distribution. The gluon densities obtained with $\mu _{s}=1$~GeV and $\bar{q}_{s}=1.4$~GeV, and
SaS input distribution with $\mu _{s}=\bar{q}_{s}=1.4$~GeV showed the
largest differences compared to the gluon density obtained with the
default option (see Figure~\ref{uncertain}). Therefore, the cross sections were
calculated also with these gluon densities to get an estimate of the error. 

\textbf{The strong coupling constant $\alpha _{s}$} is, despite its name,
not a constant, but varies with the energy scale $\mu$, according to
(in a first approximation)
\begin{equation}
\alpha _{s} = \frac{12\pi}{(33-2\cdot n_{f})\ln (\frac{\mu ^{2}}{\Lambda ^{2}})}.
\label{alphas}
\end{equation}
Here, $n_{f}$ is the number of quark flavours with mass less than the energy
scale $\mu$, and $\Lambda$ is a constant which has to be determined
by experiments. The value of $\Lambda$ also depends on the order that
the process is calculated at. However, it is not clear how to compare
the value of $\Lambda$ used in the CCFM approach with the one used in
NLO DGLAP, since the first
order in the CCFM approach includes some second order diagrams in the
DGLAP approach. The default value was chosen to be $\Lambda =
0.2$~GeV, which was also used in the evolution of the unintegrated
gluon density, but
$\Lambda = 0.34$~GeV (used in~\cite{Drees}) was also tested.  

\textbf{The energy scale $\mu$} is of the order of the typical momentum
transfer, but the exact choice of scale should be such that the higher
order contributions in the perturbative expansion of the cross section
are minimal. The standard choice is $\mu^{2}=m^{2}+p_{t}^{2}$, where
$m$ is the heavy quark mass and $p_{t}$ is the transverse momentum of
the heavy quarks in the center of mass frame of the heavy quark system. To show
the effect of the choice of $\mu$, the scale $\mu^{2}=4m^{2}$ is also
used here as a comparison.

\textbf{The quark masses.} Since the quarks are always confined in
colorless hadrons, their masses cannot be measured directly. Instead,
the masses must be extracted from hadron properties. However, this can
be done in various ways depending on the exact definition of the quark
masses, and different methods lead to slightly different
results. Therefore, the heavy quark masses are varied such that
$4.5$~GeV $< m_{b} < 5$~GeV and $1.3$~GeV $< m_{c} < 1.7$~GeV. This
variation of the quark masses gives the largest uncertainty, and can
therefore be used as an estimate of the total uncertainty, as seen in
Figure~\ref{xsec2norm}, for example.

In Figure~\ref{xsecvar} is shown the cross sections for heavy quark
production in $e^{+}e^{-}$ collisions, calculated with different
choices of $\Lambda$, $\mu ^{2}$ and unintegrated gluon densities. One
can see that a change in $\Lambda$ from $\Lambda=0.2$~GeV to
$\Lambda=0.34$~GeV gives the largest variation of the cross sections,
while changing $\mu ^{2}$ and the gluon density has a smaller effect. However, all these uncertainties are small compared to a change
in the quark masses, see Figure~\ref{eexsec}. 

\section{Conclusions}

The unintegrated gluon density for the photon was obtained with the
full CCFM evolution for the first time. The CCFM evolved gluon density
for the photon was used as input in the MC generator CASCADE, and
cross sections for heavy quark production in $e^{+}e^{-}$ collisions were calculated and compared to LEP data. Also,
predictions for charm and beauty production in $e^{+}e^{-}$ and
$\gamma \gamma$ collisions at TESLA energies are given. 

In $e^{+}e^{-}$ collisions, the obtained cross sections were somewhat larger than NLO DGLAP
predictions, but in good agreement with the charm data
from LEP. The CCFM approach made a slight improvement compared to the
standard collinear approach
for the beauty cross sections. However, the improvement did not fully
account for the discrepancy to the $b\bar{b}$ data.


\section{Acknowledgements}
This work was supported by INTAS grant 00-00679.


\end{document}